\documentstyle[12pt,equation]{article}

\usepackage{equations}

\begin{document}
\newcommand{\mlsp}{\mbox{$m_{\tilde{Z}_1}$}}
\newcommand{\hone}{\mbox{$\langle H_1^0 \rangle$}}
\newcommand{\htwo}{\mbox{$\langle H_2^0 \rangle$}}
\newcommand{\tanb}{\mbox{$\tan \! \beta$}}
\newcommand{\ml}{\mbox{$m_{H_l}$}}
\newcommand{\matp}{\mbox{${\cal M}_p ( \chi \chi \to$}}
\newcommand{\mats}{\mbox{${\cal M}_s ( \chi \chi \to$}}
\newcommand{\pprop}{\frac{m^2_{\chi}}{4 m^2_{\chi} - m_P^2 + i m_P \Gamma_P}}
\newcommand{\zprop}{\frac{m^2_{\chi}}{4 m^2_{\chi} - M_Z^2 + i M_Z \Gamma_Z}}
\newcommand{\hprop}{\frac{m^2_{\chi}}{4 m^2_{\chi} - M_{H_i}^2 + i M_{H_i}
\Gamma_{H_i}}}
\newcommand{\pro}{\mbox{$M_1 \cdot \mu \cdot \tan \! \beta$}}
\newcommand{\oh}{\mbox{$\Omega h^2$}}
\newcommand{\ttbar}{\mbox{$t \overline{t}$}}
\newcommand{\ffbar}{\mbox{$f \overline{f}$}}
\newcommand{\cb}{\mbox{$\cos \! \beta$}}
\newcommand{\sib}{\mbox{$\sin \! \beta$}}
\newcommand{\cw}{\mbox{$\cos \! \theta_W$}}
\newcommand{\sw}{\mbox{$\sin \! \theta_W$}}
\newcommand{\mlsq}{\mbox{$m^2_{H_l}$}}
\newcommand{\mpl}{\mbox{$M_P$}}
\newcommand{\mx}{\mbox{$M_X$}}
\newcommand{\msq}{\mbox{$m_{\tilde{q}}$}}
\newcommand{\msqsq}{\mbox{$m^2_{\tilde{q}}$}}
\newcommand{\msb}{\mbox{$m^2_{\tilde{b}_1}$}}
\newcommand{\msbb}{\mbox{$m^2_{\tilde{b}_2}$}}
\newcommand{\mstau}{\mbox{$m_{\tilde{\tau}_1}$}}
\newcommand{\mst}{\mbox{$m^2_{\tilde{t}_1}$}}
\newcommand{\mstt}{\mbox{$m^2_{\tilde{t}_2}$}}
\newcommand{\msn}{\mbox{$m_{\tilde{\nu}}$}}
\newcommand{\mlr}{\mbox{$m_{\tilde {l}_R}$}}
\newcommand{\msnsq}{\mbox{$m^2_{\tilde{\nu}}$}}
\newcommand{\sig}{\mbox{$\sigma_{\rm ann}$}}
\newcommand{\gstar}{\mbox{$g_{*}$}}
\newcommand{\evec}{\vec{e}}
\newcommand{\xf}{\mbox{$x_F$}}
\newcommand{\mchi}{\mbox{$m_{\chi}$}}
\newcommand{\Omchi}{\mbox{$\Omega_{\chi}$}}
\newcommand{\aem}{\mbox{$\alpha_{em}$}}
\newcommand{\as}{\mbox{$\alpha_s$}}
\newcommand{\be}{\begin{equation}}
\newcommand{\ee}{\end{equation}}
\newcommand{\een}{\end{subequations}}
\newcommand{\ben}{\begin{subequations}}
\newcommand{\beq}{\begin{eqalignno}}
\newcommand{\eeq}{\end{eqalignno}}
\renewcommand{\thefootnote}{\fnsymbol{footnote} }
\hyphenation{mi-ni-mized}
\pagestyle{empty}
\noindent
\begin{flushright}
DESY 92--101\\
SLAC PUB-5860\\
July 1992
\end{flushright}
\vspace{1.5cm}
\begin{center}
{\Large \bf The Neutralino Relic Density in Minimal $N=1$ Supergravity} \\
\vspace{5mm}
Manuel Drees\\
{\em Theorie-Gruppe, DESY, Notkestr. 85, D2000 Hamburg 52, Germany} \\
\vspace{5mm}
Mihoko M. Nojiri\footnote{Nishina Fellow. Work supported in part by
the Department of Energy, Contract DE-AC03-76SF00515, and
 the Grant--in--Aid for Scientific
 Research from the Ministry of Education,
Science and Culture, no. 02952050. E-mail: KEKVAX::NOJIRIN, NOJIRIM@SLACVM}\\
{\em Stanford Lenear Accelerator Center, Stanford University,CA94309,USA}
\end{center}
\begin{abstract}

{\small We compute the cosmic relic (dark matter) density of the lightest
supersymmetric particle (LSP) in the framework of minimal $N=1$ Supergravity
models with radiative breaking of the electroweak gauge symmetry. To this end,
we re--calculate the cross sections for all possible annihilation processes
for a general, mixed neutralino state with arbitrary mass. Our analysis
includes effects of all Yukawa couplings of third generation fermions, and
allows for a fairly general set of soft SUSY breaking parameters at the Planck
scale. We find that a cosmologically interesting relic density emerges
naturally over wide regions of parameter space. However, the requirement that
relic neutralinos do not overclose the universe does not lead to upper bounds
on SUSY breaking parameters that are strictly valid for all combinations of
parameters and of interest for existing or planned collider experiments; in
particular, gluino and squark masses in excess of 5 TeV cannot strictly be
excluded. On the other hand, in the ``generic'' case of a gaugino--like
neutralino whose annihilation cross sections are not ``accidentally'' enhanced
by a nearby Higgs or $Z$ pole, all sparticles should lie within the reach of
the proposed $pp$ and $e^+e^-$ supercolliders. We also find that requiring the
LSP to provide all dark matter predicted by inflationary models imposes a
strict lower bound of 40 GeV on the common scalar mass $m$ at the Planck scale,
while the lightest sleptons would have to be heavier than 100 GeV. Fortunately,
a large relic neutralino density does not exclude the possibility that
charginos, neutralinos, gluinos and squarks are all within the reach of
LEP200 and the Tevatron.}
\end{abstract}
\clearpage
\noindent
\pagestyle{plain}
\setcounter{page}{1}
\section*{1. Introduction}
It is by now well established \cite{1} that the observed, luminous matter in
the universe cannot account for its total mass. Cosmological mass densities are
usually expressed as ratio $\Omega \equiv \rho / \rho_c$, where $\rho_c \simeq
2 \cdot 10^{-29} h^2 {\rm g/cm^3}$ is the ``critical'' mass density that yields
a flat universe, as favoured by inflationary cosmology \cite{2}; $\rho < \ (>)
\ \rho_c$ corresponds to an open (closed) universe, i.e. a metric with negative
(positive) curvature. The dimensionless parameter $h$ is proportional to the
Hubble ``constant'' $H$ describing the expansion of the universe: $H \equiv 100
h$ km/sec Mpc. Observations yield $0.5 \leq h \leq 1$.
Even if one broadly defines luminous matter as
everything that emits any kind of electromagnetic radiation, when averaged over
the volume of the (visible) universe it cannot give $\Omega > 0.01$. In
contrast, from the observed orbits of hydrogen clouds around a variety of
galaxies, including our own, one derives \cite{1} $\Omega \geq 0.1$; and from
the motion of (clusters of) galaxies within superclusters one can deduce
\cite{1} $\Omega \geq 0.3 - 0.4$. Finally, as indicated above, $\Omega = 1$ is
predicted \cite{2} by models of inflationary cosmology; such models are
currently favoured, since they can solve other cosmological problems, e.g. the
flatness, horizon and magnetic monopole problems.

The nature of the missing or dark matter (DM) cannot be derived directly from
present observations. However, within standard (Big Bang) cosmology, the
observed abundances of light elements (D, He, Li) can only be understood
\cite{3} if the total baryonic mass density is less than about 0.1 $\rho_c$.
The discrepancy between the lower bound on the total mass density and the upper
bound on the baryonic one has led to speculations \cite{4} that some neutral,
weakly interacting stable particle might provide the bulk of the mass density
today. In particular, it has been observed almost 10 years ago \cite{5} that
the lightest supersymmetric (SUSY) particle (LSP) is a good candidate for dark
matter. Its stability is guaranteed by a discrete symmetry called R parity,
which can be imposed in most phenomenologically viable SUSY models; in
particular, R parity is automatically conserved in the simplest realistic
SUSY model \cite{6,7}, the minimal supersymmetric standard model or MSSM. This
explanation is especially attractive since the primary motivation for SUSY has
nothing to do with the DM problem; rather, SUSY {\em automatically} provides a
DM candidate ``for free''. Moreover, for ``natural'' choices of parameters (to
be specified below), $\Omega$ turns out to be \cite{5} of approximately the
right order of magnitude.

The primary motivation for the introduction of supersymmetry stems from the
observation \cite{8} that in SUSY models, large hierarchies between mass scales
are automatically protected against (quadratically divergent) radiative
corrections, in contrast \cite{9} to non--supersymmetric models. In particular,
within the non--supersymmetric standard model (SM) it is extremely unnatural to
assume the scale of electroweak symmetry breaking, characterized by the vacuum
expectation value (vev) of the Higgs field $\langle H \rangle = 175$ GeV, to be
much smaller than the scale of Grand Unification $\mx \simeq 10^{15} -
10^{16}$ GeV or the Planck scale $\mpl \simeq 10^{19}$ GeV. On the other hand,
in the MSSM radiative corrections are under control, provided only that the
mass scale of the superpartners is not much bigger than 1 TeV. The relation
between the scale of electroweak symmetry breaking and sparticle masses is even
more direct in minimal supergravity (SUGRA) models where  the breakdown of
electroweak gauge symmetry is induced by (logarithmic) radiative corrections to
the parameters of the Higgs potential \cite{10,7}. These models also have the
practical advantage that they allow to describe the whole spectrum of
superparticles (sparticles) in terms of a small number of free parameters.
Moreover, the resulting sparticle spectra almost automatically satisfy
constraints from $K$ and $B$ physics \cite{11}, and lead to small additional
contributions to electroweak observables \cite{12}, in agreement with LEP
results \cite{13}. Finally, precision measurements at LEP have shown \cite{14}
that the non--supersymmetric SM does not lead to a Grand Unification of the
gauge couplings, whereas in the MSSM all three gauge couplings meet at scale
$\mx \simeq 10^{16}$ GeV. While this result does not depend \cite{15,16} on
the constraints on the sparticle spectrum imposed by minimal SUGRA, it does
lend
credence to the assumption that there is no additional threshold between the
weak scale and \mx; this assumption is an important ingredient of SUGRA models
with radiative symmetry breaking.

Quite a few papers have already been published in the last decade that contain
calculations of the relic LSP density in some version of the MSSM. However,
older papers \cite{5,17} often assume rather light sparticles, in conflict with
recent experimental bounds. Moreover, many previous calculations
\cite{5,17,18,19,19a} involved simplifying assumptions about the sparticle
spectrum, which were often in conflict with SUGRA predictions (e.g. by assuming
the LSP to have the same mass as squarks). Some computations \cite{20,21,22,23}
do take the SUGRA relations between sfermion and gaugino masses into account,
but the additional constraints on model parameters imposed by radiative gauge
symmetry breaking have still been ignored in these papers. Moreover, in
refs.\cite{20,21,23} only the case of a ``light'' LSP, with mass below that of
the $W$ boson, has been treated. Indeed, only two calculations of the
annihilations cross sections of a heavy LSP exist to date \cite{18,19}, and
neither of them is fully complete. In ref.\cite{18} the annihilation into one
Higgs boson and one gauge boson has not been included; moreover, the given
expressions do not seem to be applicable if the mass of the LSP is less than
half the mass of the heaviest Higgs boson. Ref.\cite{19} does treat the full
list of possible final states, but only considers unmixed (pure) neutralino
states; we will see that this actually gives wrong results for the
gaugino--like LSP even in the limit of infinite LSP mass. The results of
ref.\cite{19} have been used in refs.\cite{19a,22,24} as well. We have
independently computed {\em all} relevant annihilation cross sections for a
{\em general}, mixed LSP eigenstate. This part of our work should be useful
beyond the context of the more restrictive SUGRA models. Very recently another
calculation of the relic density of heavy LSPs has appeared \cite{24a}. The
authors state that they generalized the results of ref.\cite{19} by including
neutralino mixing, but no explicit expressions are given; moreover, no SUGRA
mass relations are assumed.

We are aware of only two calculations of relic LSP densities \cite{25,24} in
which the constraints imposed by radiative gauge symmetry breaking have been
taken into account. However, in these papers a specific SUSY breaking scheme,
the ``no--scale'' ansatz \cite{26}, has been assumed; in this scheme the LSP
density turned out to be too low to account for all dark matter.\footnote{This
result can also be derived from refs.\cite{21,22} once one makes use of the
fact that these models cannot support a higgsino--like LSP; see also
ref.\cite{23}.} Moreover, the Higgs sector has only been treated in the
tree--level approximation. The importance of Higgs exchange contributions has
been pointed out in ref.\cite{27}, again using tree--level formulae for Higgs
boson masses. However, radiative corrections to these masses can be \cite{28}
substantial if the top quark is heavy, $m_t >$ 100 GeV, which now seems
likely.\footnote{Leading one--loop corrections to the Higgs sector are included
in ref.\cite{23}, but only for light LSPs, and without the constraints
imposed by radiative gauge symmetry breaking.}

In this paper we compute the LSP relic density in minimal SUGRA models with
radiative gauge symmetry breaking. We use 3 free parameters to describe SUSY
breaking, as opposed to only 1 in refs.\cite{25,24}. Due to the constraints
imposed by radiative gauge symmetry breaking, which in particular fix the Higgs
spectrum for a given set of SUSY breaking parameters and given $m_t$, the
resulting parameter space is still sufficiently small to allow for an
exhaustive scan; we thus do not have to rely on ``simplifying assumptions'' of
often dubious validity. As indicated above, we always include neutralino
mixing, as well as {\em all} kinematically accessible final states, when
computing the LSP annihilation cross section. Our analysis of radiative
symmetry breaking includes effects of the Yukawa couplings of the $b$ quark and
$\tau$ lepton, which can be quite important \cite{32a,DN1}. Mixing between the
superpartners of left and right handed fermions is also treated exactly.
One--loop corrections to the Higgs sector are included, and all experimental
constraints on sparticle masses are taken into account.

We find that the model can easily yield sufficient dark matter to close the
universe, {\em if} the SUSY breaking common scalar mass $m$ at scale \mx\
exceeds 40 GeV. This conclusion is closely related to the result of
refs.\cite{29,30} that a single light slepton suffices to make the relic
density of a gaugino--like LSP uninterestingly small. A qualitatively similar
result has been found in ref.\cite{23} in a more general context. On the other
hand, requiring the LSP not to overclose the universe does {\em not} lead to
upper bounds on sparticle masses which are both relevant for existing or
planned experiments and are valid for the entire parameter space. Of course,
it is quite unnatural \cite{31,16} to assume
sparticles to be much heavier than the $W$ and $Z$ bosons, but naturalness
arguments cannot be translated into strict upper bounds; even a rather high
\cite{18,19} bound from cosmology would therefore have been welcome. Alas,
there are two different ways in which such a bound can be circumvented. One
possibility is to have a light higgsino--like LSP, with all SUSY breaking
masses being very large. Note that the higgsino mass does {\em not} break
supersymmetry; moreover, we argue that in such a scenario, supersymmetry would
be extremely difficult to discover in laboratory experiments. The constraints
from radiative gauge symmetry breaking exclude the possibility to have a
higgsino--like LSP if the top quark is heavy, the precise bound on $m_t$
depending on the ratios of the soft SUSY breaking parameters. The
second possibility to allow for a very heavy sparticle spectrum is to choose
the LSP mass to be close to half the mass of the pseudoscalar Higgs boson; this
strongly enhances the annihilation of the LSP into SM fermions, in particular
$b$ quarks and $\tau$ leptons. Within the framework of minimal SUGRA, this
scenario necessitates a large ratio \tanb\ of the vevs of the two Higgs fields
of the model, but such solutions can be realized quite easily \cite{32a,DN1}.

The remainder of this paper is organized as follows. In sec. 2 we briefly
describe the formalism necessary to compute the DM density for a given LSP
annihilation cross section. We also list all possible LSP annihilation
processes, and give a short description of the parameter space of minimal
SUGRA models. In sec. 3 we present illustrative examples of LSP densities
in such models. The effects of $s$--channel poles as well as thresholds,
where new annihilation channels open up, are discussed. We also explicitly
demonstrate the importance of the SUGRA imposed running of sfermion and Higgs
boson masses. Finally, we study the dependence of the DM density on the free
parameters of the model by means of several contour plots. Sec. 4 is devoted to
a discussion of the bounds that can be derived from the requirement that the
LSP relic density lies in the cosmologically interesting region. As already
mentioned above, a strict upper bound on sparticle masses can only be derived
if we artificially restrict the parameter space of the model. In contrast,
a non--trivial {\em lower} bound on slepton masses can be derived from the
requirement that the LSP relic density be close to the critical density;
however, the masses of the gluino, the light chargino, and the squarks
can all be near their present lower bounds. Finally, in sec. 5
we summarize our results and draw some conclusions. Complete lists of all LSP
annihilation matrix elements are given in Appendix A, and Appendix B contains
an example of a check of some amplitudes for the production of longitudinal
gauge bosons using the equivalence theorem.
\setcounter{footnote}{0}
\section*{2. Formalism}
In this section we describe the formalism necessary to describe the numerical
results of secs. 3 and 4. We first (sec. 2a) give a short summary of the
calculation of the present day DM density for given mass and annihilation cross
section of the dark matter candidate $\chi$. Sec. 2b contains a brief
description of the MSSM; it also contains a list of all annihilation channels
of our DM candidate. The resulting annihilation cross section will turn out
to depend on the {\em whole} sparticle and Higgs boson spectrum. In sec. 2c we
therefore give a brief summary of minimal SUGRA models, where the whole
spectrum
can be computed in terms of 4 free parameters.

\subsection*{2a. Calculation of the DM density}
We begin with a brief description of the calculation of the present relic mass
density of a DM candidate $\chi$, assuming that the mass \mchi\ as well as
the annihilation cross section $\sig (\chi \chi \to {\rm anything})$ are known.
Following the prescription of refs.\cite{4,18}, we first introduce the
freeze--out temperature $T_F$, below which the $\chi \chi$ reaction rate is
(much) smaller than the expansion rate of the universe. It is convenient to
express (inverse) temperatures in terms of the dimensionless quantity
$x \equiv \mchi/T$; the freeze--out temperature can then iteratively be
computed from \be \label{e1}
\xf = \ln \frac { 0.0764 \mpl \left( a + 6 b / \xf \right) c ( 2 + c) \mchi}
{ \sqrt{ \gstar \xf} }. \ee
Here, $\mpl = 1.22 \cdot 10^{19}$ GeV is the Planck mass, and $a$ and $b$ are
the first two coefficients in the Taylor expansion of the annihilation cross
section with respect to the relative velocity $v$ of the $\chi \chi$ pair in
its
center--of--mass frame: \be \label{e2}
v \cdot \sig (\chi \chi \to {\rm anything}) = a + b v^2; \ee
notice that $v$ is {\em twice} the velocity of $\chi$ in the $\chi \chi$ cms
frame. Furthermore, \gstar\ is the effective number of relativistic degrees of
freedom at $T = T_F$. A highly relativistic boson (fermion) contributes 1 (7/8)
to \gstar, whereas very nonrelativistic (slow) particles do not contribute at
all. However, often $T_F$ turns out to be close to the mass of the heavy
particles of the SM (the $c$, $b$ or $t$ quark, the $\tau$ lepton, or the $W$
or $Z$ boson); in this case a careful treatment of the threshold is necessary
\cite{4} if ugly jumps in the curves are to be avoided. Finally, the constant
$c$ in eq.(\ref{e1}) is a numerical parameter introduced \cite{4} to achieve
smooth matching of approximate solutions of the Boltzmann equation above and
below $T_F$; following ref.\cite{18} we chose $c = 1/2$. Given \xf\ and
\sig, one can compute \be \label{e3}
\Omchi h^2 \equiv \frac {\rho_{\chi}} {\rho_c / h^2}
 = \frac {1.07 \cdot 10^9 / {\rm GeV} \xf} {\sqrt{\gstar} \mpl \left( a +
3 b / \xf \right) }, \ee
where the rescaled Hubble constant $h$ and the critical density $\rho_c$
have already been introduced in sec. 1.\footnote{Within the framework of
inflationary cosmology, eq.(\ref{e3}) should strictly speaking be
interpreted as a calculation of the expansion rate, rather than a calculation
of the density. After all, inflation predicts \cite{2} $\Omega = 1$ to high
precision, {\em independent} of the details of the particle physics model.
However, only for a small range of values of the (absolute) mass density $\rho$
does $H$, and thus the age of the universe, come out close to the observed
value.}

Eqs.(\ref{e1})--(\ref{e3}) describe a simple, approximate solution of the
Boltzmann equation that determines the abundance of any particle species. While
not strictly correct \cite{32}, in most cases this treatment reproduces the
exact numerical solution to 10--20 \% accuracy \cite{4,18}; given that $h^2$
in eq.(\ref{e3}) is only known to a factor of 2, this accuracy is fully
sufficient for our purposes. However, it has recently been pointed out
\cite{33}
that there are three cases in which this approximation fails badly: close to a
threshold where a new annihilation channel opens up that dominates the total
annihilation cross section; close to a very narrow $s-$channel resonance; and
if the next--to--lightest sparticle $\chi'$ is close in mass to the LSP, and
$\sig (\chi \chi' \to {\rm anything}) \gg \sig (\chi \chi \to {\rm anything})$.
Unfortunately, in these three cases the proper treatment \cite{33} is
considerable more cumbersome than eqs.(\ref{e1})--(\ref{e3}). For reasons of
computational simplicity we will therefore use the approximate treatment
throughout, but we will be careful to point out the situations where it
might fail, and will qualitatively describe the result of the proper treatment
in such cases.

Note that \xf\ in eq.(\ref{e3}) is almost independent of \mchi\ and \sig; we
find $15 \leq \xf \leq 30$ for experimentally allowed choices of parameters
(see below). The number of degrees of freedom \gstar\ at $T_F \simeq \mchi/20$
increases monotonically with \mchi, but again the \mchi\ dependence is quite
small: $8 \leq \sqrt{\gstar} \leq 10$ for 20 GeV $\leq \mchi \leq$ 1 TeV.
The details of the model (particle masses and couplings) affect the prediction
for \Omchi\ therefore predominantly through the coefficients $a$ and $b$
describing the annihilation cross section; we now briefly describe the
calculation of this cross section.

\setcounter{footnote}{0}
\subsection*{2b. The annihilation cross section}
Within the MSSM, only the lightest neutralino is left \cite{34} as a viable DM
candidate. A stable LSP has to be \cite{35} both electrically and color
neutral,
since otherwise it would have been found in searches for exotic isotopes. This
leaves one with the lightest neutralino and the lightest sneutrino, but the
latter is excluded by a combination of the bound on the invisible decay width
of the $Z$ boson obtained at LEP \cite{36} and the limits derived from the
unsuccessful search for relic sneutrinos using Germanium detectors \cite{37}.

The sparticle spectrum of the MSSM contains \cite{6,7} 4 neutralino states:
The superpartners of the $B$ and $W_3$ gauge bosons, and the superpartners of
the neutral Higgs bosons $H_1^0$ and $H_2^0$ with hypercharge -1/2 and +1/2,
respectively. However, after electroweak gauge symmetry breaking these current
eigenstates mix; their mass matrix in the basis ($\tilde{B}, \tilde{W}_3,
\tilde{h}_1^0, \tilde{h}_2^0)$ is given by \be \label{e4}
{\cal M}^0 = \mbox{$ \left( \begin{array}{cccc}
M_1 & 0 & -M_Z \sw \cb & M_Z \sw \sib \nonumber \\
0 & M_2 & M_Z \cw \cb & -M_Z \cw \sib \nonumber \\
-M_Z \sw \cb & M_Z \cw \cb & 0 & -\mu \nonumber \\
M_Z \sw \sib & -M_Z \cw \sib & -\mu & 0 \end{array} \right), $} \ee
where we have used the convention of refs.\cite{7,38}, which we will follow
throughout. Assuming Grand Unification of the gauge couplings implies a
relation between the SUSY breaking gaugino masses $M_1$ and $M_2$ as well as
the gluino mass $m_{\tilde{g}} = |M_3|$: \be \label{e5}
M_1 = \frac {5}{3} \tan^2 \! \theta_W M_2 = \frac {5 \alpha_{em}}
{3 \alpha_s \cos^2 \! \theta_W} M_3, \ee
where \aem\ is the electromagnetic coupling constant and \as\ is the strong
coupling. The angle $\beta$ in eq.(\ref{e4}) is defined via $\tanb \equiv
\langle H_2^0 \rangle / \langle H_1^0 \rangle$. Finally, the parameter $\mu$
describes the supersymmetric contribution to the Higgs(ino) masses.

In general the LSP is a complicated mixture \cite{39,38,40} of the 4 current
eigenstates; in our numerical calculations we take full account of this
mixing by diagonalizing the mass matrix (\ref{e4}) numerically. However, in
the limit $|M_1| + |\mu| \gg M_Z$ this diagonalization can quite easily be
carried out perturbatively. Since this proves helpful for a
qualitative understanding of our numerical results, we list the eigenvalues
$m_i$ and eigenvectors $\evec_i$ of the mass matrix (\ref{e4}), keeping terms
up to ${\cal O}(M_Z)$: \ben \label{e6} \beq
m_1 &= M_1 ; \nonumber \\
\evec_1 &= \left( 1, 0, \frac{ M_Z \sw (\cb M_1 + \sib \mu)}
{\mu^2 - M_1^2}, \frac { M_Z \sw ( \sib M_1 + \cb \mu)} {M_1^2 - \mu^2}
\right);
\label{e6a} \\
m_2 &= M_2 ; \nonumber \\
\evec_2 &= \left( 0, 1, \frac{ M_Z \cw (\cb M_2 + \sib \mu)}
{M_2^2 - \mu^2}, \frac{ M_Z \cw ( \sib M_2 + \cb \mu)} {\mu^2 - M_2^2} \right);
\label{e6b} \\
m_3 &= \mu (1 + \delta); \nonumber \\
\evec_3 &= \frac {1} {\sqrt{2}} \left( \frac { M_Z \sw
(\cb+\sib) } {M_1 - \mu}, \frac{ M_Z \cw (\cb+\sib)} {\mu - M_2}, 1, -1 +
\epsilon \right) ; \label{e6c} \\
m_4 &= -\mu (1 + \delta'); \nonumber \\
\evec_4 &= \frac {1} {\sqrt{2}} \left( \frac { M_Z \sw
(\cb-\sib) } {M_1 + \mu}, \frac{ M_Z \cw (\sib+\cb)} {M_2 + \mu}, 1, 1 +
\epsilon' \right) ; \label{e6d} \eeq \een
Note that eq.(\ref{e5}) implies $|m_2| > |m_1|; \ \evec_2$ therefore never
corresponds to the LSP. In the limit where either $|M_1|$ or $|\mu|$ (or both)
is much bigger than $M_Z$, the LSP is therefore an almost pure bino
($\evec_1$), or an almost pure antisymmetric or symmetric higgsino
($\evec_{3,4}$).\footnote{Unless $|M_1| \simeq |\mu|$, in which case strong
higgsino--bino mixing occurs. Such mixed states always lead to very small relic
densities \cite{18,21,29,24a}.}

The corrections $\delta, \delta', \epsilon$ and $\epsilon'$ are formally
${\cal O}(M_Z^2)$; however, they can be numerically quite important for a
light higgsino--like LSP. They are given by \ben \label{e7} \beq
\delta &= \frac {M_Z^2 ( 1 + \sin \! 2 \beta) } { 2 \mu} \left(
\frac {\sin^2 \! \theta_W} { \mu - M_1} + \frac {\cos^2 \! \theta_W} {\mu -
M_2}
\right) ; \label{e7a} \\
\delta' &= \frac {M_Z^2 ( 1 - \sin \! 2 \beta) } { 2 \mu} \left(
\frac {\sin^2 \! \theta_W} { \mu + M_1} + \frac {\cos^2 \! \theta_W} {\mu +
M_2}
\right) ; \label{e7b} \\
\epsilon &= - \delta \frac {\cos \! 2 \beta} {1 + \sin \! 2 \beta};
\label{e7c} \\
\epsilon' &= - \delta' \frac {\cos \! 2 \beta} {1 - \sin \! 2 \beta}.
\label{e7d} \eeq \een
After inclusion of these terms one therefore finds that the mass splitting
between the two higgsino--like states is given by (in the limit $|M_1| \gg
|\mu|$): \be \label{e8}
||m_3| - |m_4|| = \frac {8 M_Z^2 \sin^2 \theta_W} {3 |M_1|} \left( 1 +
{\cal O} (\frac {\mu}{M_1}, \frac{M_Z}{M_1}) \right). \ee
More details about neutralino masses and mixings can be found, e.g., in
refs.\cite{17,38,39,40}.

What are the final states into which a pair of neutralinos can annihilate? Here
we only include 2--body final states that can be produced in leading order of
perturbation theory. From unsuccessful sparticle searches at LEP \cite{41} as
well as the bound $m_{\tilde{g}} > 120$ GeV that follows from the preliminary
gluino search limit of the CDF collaboration \cite{42} after inclusion of
``cascade decays'' \cite{43}, one can derive \cite{44} the bound $\mchi \geq
20$ GeV.\footnote{We note in passing that this implies a freeze--out
temperature $T_F \geq 1$ GeV, well above the temperature where the
quark--hadron phase transition is expected to occur. Our results do therefore
not depend on the exact value of the critical temperature for this phase
transition, in contrast to the case of a very light LSP \cite{32}.}  We see
that the annihilation $\chi \chi \to f \overline{f}$ is always kinematically
allowed for all light SM fermions up to and including $b$ quarks. For heavier
neutralinos annihilation into a pair of gauge bosons also has to be
included \cite{18,19}. In addition the model contains at least one neutral
scalar Higgs boson $h$ with mass not much above $M_Z$, even after inclusion of
radiative corrections \cite{28}; the second neutral scalar $H$, the
pseudoscalar $P$ and the charged Higgs boson $H^+$ can also be accessible. In
general one therefore also has to include annihilation into two Higgs bosons
\cite{18,19}, as well as into one Higgs and one gauge boson.

We will
now discuss annihilation into these final states in a little more detail,
assuming $\chi$ to be either a nearly pure bino or a nearly pure higgsino.
Here we only give symbolic expressions for the matrix elements, which allow to
estimate the order of magnitude of the various contributions;
exact expressions for the cross sections for a general $\chi$ state are
listed in Appendix A. Since we expand the annihilation cross section of
eq.(\ref{e2}) only up to ${\cal O}(v^2)$, we only have to include $s$ and
$p$ wave contributions. $s$ wave contributions start at ${\cal O}(v^0)$, but
also contain ${\cal O}(v^2)$ terms that contribute to eq.(\ref{e2}) via
interference with the ${\cal O}(v^0)$ terms. $p$ wave matrix elements start
at ${\cal O}(v)$, so that we only need the leading term in the expansion. Of
course, there is no interference between $s$ and $p$ wave contributions, and
hence no ${\cal O}(v)$ term in eq.(\ref{e2}). Notice finally that Fermi
statistics forces the $s$ wave state of two identical Majorana fermions
to have $CP = -1$, while the $p$ wave
has $CP = +1$; the same argument also implies that the $s$ wave has to have
total angular momentum $J=0$.

\subsubsection*{$\chi \chi \to \ffbar $}
This reaction proceeds via the $s$--channel exchange of a $Z$ or Higgs
boson, as well as via sfermion exchange in the $t-$channel.
Each chirality state of $f$ has its own superpartner with in general
different mass, which mix \cite{45} if the Yukawa coupling of $f$ is not
negligible. In the following expressions summation over both sfermions is
always understood. Note that both the $Z-f-\overline{f}$ and
fermion--sfermion--gaugino couplings conserve chirality; the sfermion and
$Z$ exchange contributions to the $s$ wave matrix element ${\cal M}_s$ are
therefore proportional to the mass $m_f$ of the final state fermions.
Contributions from Higgs exchange, from the higgsino--sfermion--fermion
Yukawa interactions, and from sfermion mixing violate chirality, but have
an explicit factor of $m_f$. The coefficient $a$ in the expansion
(\ref{e2}) of the annihilation cross section is therefore always
proportional to $m_f^2$ for this final state, independent of the composition of
$\chi$. Moreover, since the $CP$ quantum number of the exchanged Higgs boson
must match that of the initial state, only $P$ exchange contributes to
${\cal M}_s$, while $h$ and $H$ exchange contribute to ${\cal M}_p$. Since
${\cal M}_p$ only contributes to the coefficient $b$ in eq.(\ref{e2}), which is
suppressed by a factor $3/x_F \simeq 0.1 - 0.2$, $P$ exchange is potentially
much more important than the contribution from the scalar Higgs bosons.

For a bino--like LSP the matrix elements thus have the
structure \ben \label{e9} \beq
\left. \mats \ffbar )\right|_{\rm bino}
\propto g'^2 m_f & \left[  c_1 \frac {\mchi} {m^2_{\tilde f} + m^2_{\chi}}
Y_f^2 + c_2 \frac {M_Z^2} {M_1^2 - \mu^2} \frac
{m_{\chi}}{M_Z^2} \nonumber \right. \\ & \left.
+ c_3 \frac{1} {M_1 + \mu} \pprop \right]; \label{e9a} \\
\left. \matp \ffbar )\right|_{\rm bino}
\propto g'^2 v & \left[  d_1  \frac {m^2_{\chi}} {m^2_{\tilde f}+m^2_{\chi}}
Y_f^2 + d_2 \frac {M_Z^2} {M_1^2 - \mu^2}
\zprop \right. \nonumber \\ & \left.
+ \sum_{i=1}^2 d_{3,i} \frac{m_f} {M_1 + \mu} \hprop \right]. \label{e9b}
\eeq \een
Here the $c_i, \ d_i$ are numerical constants of order 1, and $g'$ is the
$U(1)_Y$ gauge coupling. $c_3$ and the $d_{3,i}$ describe the Higgs--\ffbar\
couplings, where we have introduced the notation \cite{38} $H_1 = H, \ H_2 =
h$; in certain cases some of these couplings can be enhanced \cite{38} by a
factor \tanb\ (or suppressed by $\cot \! \beta$).

We see from eqs.(\ref{e9}) that  the $s-$channel diagrams are all suppressed
by small couplings. The $Z$ boson couples to neutralinos only via their
higgsino components; more precisely, \be \label{e9'}
g_{Z \chi_i \chi_j} \propto e_{i,3} e_{j,3} - e_{i,4} e_{j,4}, \ee
where $e_{i,l}$ is the $l-$th component of $\evec_i$, see
eqs.(\ref{e6}). For a bino--like state this coupling is doubly suppressed,
as indicated in eqs.(\ref{e9}). The Higgs boson couplings to neutralinos
originate from the Higgs--higgsino--gaugino gauge interaction terms of the
unmixed Lagrangian; for a bino--like neutralino this coupling involves
therefore only one factor of the (small) higgsino component. Finally, the
sfermion exchange contributions can only be suppressed by choosing
$m^2_{\tilde f} \gg m^2_{\chi}$ in this case.

For a higgsino--like LSP, eqs.(\ref{e9}) become: \ben \label{e10} \beq
\left. \mats \ffbar)\right|_{\rm higgsino} \propto (g^2+g'^2) m_f
& \left[ \left( c_1' \frac
{M_Z} {\mu + M} + c_1'' \frac {m_f}{M_Z} \right)^2 \frac {\mchi}
{m^2_{\tilde f}+m^2_{\chi}} + c_2' \frac {M_Z^2} {\mu M} \frac {\mchi} {M^2_Z}
\nonumber \right. \\ & \left.
+ c_3' \frac{1} {M_1 + \mu} \pprop \right]; \label{e10a} \\
\left. \matp \ffbar)\right|_{\rm higgsino} \propto (g^2+g'^2) v
& \left[ \left( d_1' \frac
{M_Z} {\mu + M} + d_1'' \frac {m_f}{M_Z} \right)^2 \frac {m^2_{\chi}}
{m^2_{\tilde f}+m^2_{\chi}} \nonumber \right. \\ & \left.
+ c_2' \frac {M_Z^2} {\mu M} \zprop
\nonumber \right. \\ & \left.
+ \sum_{i=1}^2 d_{3,i}' \frac{m_f} {M_1 + \mu} \pprop \right]; \label{e10b}
 \eeq \een
Notice that eqs.(\ref{e10}) also get contributions
from $SU(2)$ gauge interactions,
which enter eqs.(\ref{e9}) only in higher orders in $M_Z/(M+\mu)$. On the
other hand, the sfermion exchange contribution is now suppressed by either
the small gaugino component of the LSP, see eqs.(\ref{e6c}),(\ref{e6d}), or by
a
power of the Yukawa coupling; of course, for $f = t$ the latter is hardly a
suppression, and can even be an enhancement if the top quark is heavy. In
contrast, Higgs exchange contributes at the same order (in $\frac {M_Z} {M_1 +
\mu}$) to the annihilation of bino--like and higgsino--like LSPs.\footnote{The
fact that the Higgs coupling to a mixed neutralino is unsuppressed explains to
a large part the big annihilation cross sections, and hence small relic
densities, of this kind of LSP.} Finally, the $Z$ exchange contribution again
behaves quite differently in the two cases of eqs.(\ref{e9}) and (\ref{e10}):
While in the former case, it decreases quadratically with the mass of the
heavy neutralinos (whose mass is $\propto \mu$, in this case),
eqs.(\ref{e7}) and (\ref{e9'}) show that the $Z \chi \chi$ coupling for a
higgsino--like LSP decreases only linearly with the mass of the heavy
neutralinos (with mass $\propto M$).\footnote{Unitarity implies that the
$Z \chi \chi$ coupling must be suppressed at least like $1/\mchi$ as $\mchi
\to \infty$. The $Z$ exchange contribution to ${\cal M}_s$ behaves like
$g_{Z \chi \chi} m_f \mchi /M_Z^2$. This contribution cannot be
cancelled by $\tilde{f}$ or $P$ exchange because one can always choose
$m_{\tilde f}, m_P \gg \mchi$; since this does not increase any couplings,
the heavy states simply decouple in this limit. The $Z$ coupling itself
therefore has to compensate for the factor of \mchi.}
Moreover, eqs.(\ref{e6}) and (\ref{e9'})
imply that the off--diagonal $Z \chi \chi'$ coupling is not suppressed at all
if $\chi = \evec_3, \ \chi' = \evec_4$ or vice versa. If $Z$ exchange gives the
dominant contribution (which is true for light higgsinos, except in the
vicinity of the $h$ pole), one therefore has \be \label{e11}
\sigma (\chi \chi' \to \ffbar) \propto \left( \frac {M_1 \mu} {M_Z^2}
\right)^2 \cdot \sigma ( \chi \chi \to \ffbar); \ee
In this case $\chi \chi'$ co--annihilation \cite{33} can be quite important
\cite{24a}; we will come back to this point later.

\subsubsection*{$\chi \chi \to W^+W^-, \ ZZ$}
These final states can be produced via $t-$channel chargino or neutralino
exchange, as well as $s-$channel exchange of scalar Higgs bosons; in case of
the $W^+W^-$ final state $s-$channel $Z$ exchange also contributes. It is
important to realize here that the cross sections behave quite differently for
longitudinal and transverse gauge bosons. For each longitudinal gauge boson the
amplitude gets an enhancement factor $\gamma_V = E_V/M_V \simeq \mchi/M_V$.
Unitarity then requires strong cancellations between different contributions
to the matrix element if the couplings of $\chi$ to $V$ are not suppressed. On
the other hand, these enhancement factors can give finite matrix elements in
the limit $\mchi \to \infty$ even if the couplings do vanish in this limit.
These effects can also be understood from the equivalence theorem \cite{46},
which states that in the high energy limit, the matrix element for the
production of a longitudinal gauge boson is identical to the one for the
production of the would--be Goldstone boson that gets ``eaten'' when the gauge
boson acquires its mass. The couplings of these Goldstone modes to neutralinos
and charginos originate from the Higgs--higgsino--gaugino gauge interactions of
the unmixed Lagrangian.
This means that a pair of would--be Goldstone bosons can be produced
with full gauge strength from pure gaugino as well as pure higgsino initial
states, by exchange of higgsinos or gauginos, respectively. On the other
hand, since the relevant couplings are gauge couplings, the matrix element
must be well behaved in the limit where any mass becomes very large.
Notice finally that neither a pair of longitudinal
gauge bosons nor a combination of one longitudinal and one transverse gauge
bsoson can exist in a $J=0$ state with $CP=-1$; these final states are
therefore
only accessible to the $p-$wave initial state.

For a
bino--like LSP, the coupling to gauge bosons does indeed decrease like
$1/\mchi$, see eq.(\ref{e6a}). The above discussion then shows that only
the amplitude for the production of two longitudinal gauge bosons survives,
which is purely $p-$wave: ($V = W, \ Z$): \be \label{e12}
\left. \matp VV)\right|_{\rm bino} \propto g'^2 v \left[ d_4 \frac {M_Z^2}
{M_1^2 + \mu^2} \frac {\mchi} {\mu} + \sum_{i=1}^2 d_{5,i}
\frac {M_Z} {M_1 + \mu}
\frac {\mchi M_V} {4 m_{\chi}^2 - m^2_{H_i} + i m_{H_i} \Gamma_{H_i}}
\right] \frac {m^2_{\chi}}{M^2_V}. \ee
Because of the enhancement factor $\gamma_V^2$, the annihilation of heavy
bino--like LSPs into (longitudinal) gauge bosons does {\em not} vanish for
$\mchi \to \infty$, unless one has $|M_1| \ll |\mu|$. Considering \cite{19} an
exact bino state therefore does not give the right answer in this case. As
discussed above, this can also be understood from the equivalence theorem,
since
the production of two Goldstone modes can only be suppressed by making the
exchanged higgsino very heavy.  (This argument is made more rigorous in
Appendix B, where numerical factors from the diagonalization of the Higgs
sector
etc. are treated properly.) Finally, we mention that the coefficient $d_{5,1}$
is often quite small, since the heavy Higgs scalar decouples \cite{38} from $W$
and $Z$ bosons when its mass is large. In contrast the exchange of the light
Higgs boson actually gives the dominant contribution in the limit $|\mu| \gg
|M_1|$.

For a higgsino--like LSP one has \ben \label{e13} \beq
\left. \mats VV)\right|_{\rm higgsino} &\propto (g^2+g'^2) c_4'; \label{e13a}
\\
\left. \matp VV)\right|_{\rm higgsino} &\propto (g^2+g'^2) v \left[ d_4'
\right. \nonumber \\ & \left. \hspace*{1cm}
+ \sum_{i=1}^2 d_{5,i}' \frac {M_Z} {M + \mu}
\frac {\mchi M_V} {4 m_{\chi}^2 - m^2_{H_i} + i m_{H_i} \Gamma_{H_i}}
\frac {m^2_{\chi}}{M^2_V} \right]. \label{e13b} \eeq \een
Notice that in this case there is no propagator suppression of the $t-$channel
diagrams for $V_TV_T$ production, because the exchanged (other) higgsino state
is almost mass degenerate with the LSP. Since the (off--diagonal) couplings to
gauge bosons are not suppressed in this case, cancellations between $t-$channel
(and $Z$ exchange, for $V=W$) diagrams are necessary to restore unitarity for
$V_LV_L$ production. The equivalence theorem shows that in the limit $|M_1| \gg
|\mu|$ the contribution from longitudinal gauge bosons is even suppressed,
since the production of Goldstone bosons necessitates the exchange of heavy
gaugino--like states, or the exchange of $Z$ or Higgs bosons whose diagonal
couplings are also suppressed in this limit. Moreover, the production of
$V_LV_T$ final states is suppressed by neutralino mixing factors. However, the
production of transverse gauge bosons, and hence the total cross section, is
not suppressed in the limit where the masses of the heavier neutralinos becomes
very large; this is again in contrast to the case of a bino--like
LSP.\footnote{The $V_LV_L$ final states have not been treated properly in
ref.\cite{19}. However, since they only contribute to $p-$wave annihilation,
they are numerically not very important. This is even true for bino--like
LSP's where the production of transverse gauge bosons does not contribute,
since here the total annihilation cross section is dominated by the \ffbar\
final state.}

\subsubsection*{$\chi \chi \to Z h$}
This final state can be produced via neutralino exchange in the $t-$channel as
well as $s-$channel exchange of $Z$ and $P$ bosons. Notice that now a
longitudinal $Z$ boson can be produced both from the $s-$ and
$p-$wave; since we always need at least one neutralino mixing factor, the
matrix
element would go to zero in the limit of large LSP mass without the enhancement
factor $\gamma_Z$. For the case of a bino--like
LSP, the matrix elements have the form: \ben \label{e14} \beq
\left. \mats Z h) \right|_{\rm bino} \propto g'^2 &
\frac {\mchi} {M_1 + \mu} \frac {M_Z^2} {m_P^2 + M_Z^2}
\left[ c_6   
+ c_7 \pprop \right]; \label{e14a} \\
\left. \matp Z h) \right|_{\rm bino} \propto g'^2 & v d_6 \frac {m_{\chi}^2}
{M_1^2 + \mu^2}. \label{e14b} \eeq \een
Both $c_6$ and $d_6$ get contributions from neutralino as well as $Z$ exchange
diagrams; in the important limit $m^2_P \gg M_Z^2$, ${\cal M}_s$ is strongly
suppressed, due to a cancellation between the two classes of diagrams. In this
limit the $ZhP$ coupling also becomes small \cite{38}, suppressing the $P$
exchange contribution as indicated. Only the $p-$wave amplitude (\ref{e14b})
therefore survives in the limit $m_P^2 \gg M_Z^2$. Notice, however, that the
amplitude as a whole does not vanish in the limit of large sparticle masses
and small neutralino mixing, unless $|\mu| \gg |M_1|$; once again this is due
to the production of a longitudinal gauge boson, giving rise to an enhancement
factor $\mchi/M_Z$.

For a higgsino--like LSP eqs.(\ref{e14}) become: \ben \label{e15} \beq
\left. \mats Zh ) \right|_{\rm higgsino} &\propto (g^2+g'^2)
\frac {\mchi} {M + \mu} \left[ c_6'  + c_7' \frac {M_Z^2} {m_P^2 + M^2_Z}
\frac {m^2_{\chi}} {4 m_{\chi}^2 -m_P^2 + i \Gamma_P m_P} \right]; \label{e15a}
\\
\left. \matp Zh ) \right|_{\rm higgsino} &\propto (g^2+g'^2) v
d_7' \frac{m^2_{\chi}} {M^2 + \mu^2}. \label{e15b} \eeq \een
Note that in this case the ${\cal O}(v^0)$ term from the $t-$channel and $Z$
exchange diagrams is {\em not} suppressed for $m_P^2 \gg M_Z^2$.
 Just as in the case
of a bino--like LSP the total amplitude is only suppressed if the heavier
neutralinos are much heavier than \mchi, i.e. if $|M_1| \gg |\mu|$ in this
case.

\subsubsection*{$ \chi \chi \to hh$}
Here only $t-$channel neutralino exchange and $s-$channel scalar Higgs
exchange diagrams contribute. Since two identical scalars cannot be in a
state with $J=0$ and $CP=-1$, annihilation can only proceed from the
$p-$wave. The amplitude thus has the general form:
\be \label{e16}
\matp hh ) \propto g'^2 v \left[ d_8 \frac {\mchi} {M + \mu}
+ d_9 \frac {M_Z^2} {M^2 - \mu^2}
+ \sum_{i=1}^2 d_{10,i} \frac {M_Z} {M + \mu} \frac {M_Z \mchi}
{4 m_{\chi}^2 - m_{H_i}^2 + i m_{H_i} \Gamma_{H_i}} \right]. \ee
This form holds for both bino--like and higgsino--like LSP's. The contribution
$\propto d_8$ comes from the exchange of the heavier neutralinos, which occurs
with full gauge strength but is suppressed by small propagators; the term
$\propto d_9$ originates from neutralino mixing. In case of a bino--like
LSP the coefficient $d_8$ is suppressed if $\tanb \gg 1$ and
$|\mu| \gg |M_1|$: \be \label{e17}
d_{8, {\rm bino}} = d_8' \frac {M_1 + \mu \sin \! 2 \beta} {\mu}. \ee
This possible additional suppression is absent for the case of a
higgsino--like LSP; in this case the amplitude also gets contributions from
$SU(2)$ gauge interactions, as do all other higgsino annihilation amplitudes.

\subsubsection*{  }
This concludes our qualitative discussion of the annihilation matrix elements
for the most important final states. In principle the heavier Higgs bosons $H,
\ P, \ H^+$ could also be produced \cite{18,19} in $\chi \chi$ annihilation,
either in pairs or in association with a gauge boson. However, we will see in
the next subsection that in minimal supergravity models these heavy states are
usually not accessible. We do therefore not discuss the relevant matrix
elements here; of course, they are included in the list of matrix elements in
Appendix A.

\setcounter{footnote}{0}
\subsection*{2c. The particle spectrum}
The basic assumption of minimal Supergravity (SUGRA) models is that
supersymmetry breaking can be described \cite{7} in terms of just three
parameters: A universal scalar mass $m$, a universal gaugino mass $M$, and
a universal parameter $A$ characterizing the strength of nonsupersymmetric
trilinear scalar interactions. If the particle spectrum is restricted to
that of the MSSM, which we always assume in this paper, one in addition has
to introduce a supersymmetric contribution $\mu$ to Higgs boson and higgsino
masses; the masses $\mu_1$ and $\mu_2$
of the two Higgs doublets are then given by \be \label{e18}
\mu_1^2 = \mu_2^2 = m^2 + \mu^2, \ee
while the Higgs mixing term $\mu_3^2$ is given by \be \label{e18'}
\mu_3^2 = (A-m) \cdot m. \ee
This simple form of the particle spectrum is assumed to emerge after
integrating out the fields of the ``hidden sector'' \cite{7} where local
supersymmetry is broken spontaneously. Since the decoupling of these fields
occurs at the Planck or GUT scale the spectrum will be this simple only at
ultra--high energies $Q \geq \mx$; in particular, eq.(\ref{e18}) will only
hold at these very high energies.

Of course, present day experiments, as well as DM annihilation, occur at much
smaller energy scales. The particle spectrum at energies of the order of the
weak or sparticle mass scale can be obtained by solving a set of coupled
renormalization group equations (RGE)\cite{10}; for given $m, \ M, \ A, \ \mu$
and Yukawa couplings $h_t, \ h_b$ (with $h_b=  h_{\tau}$ at the GUT scale),
minimal SUGRA specifies the boundary conditions at $Q = \mx$, which uniquely
determine the particle spectrum at lower energies. In this scheme leading
logarithms are automatically summed, i.e. all terms of order $\left( \alpha/\pi
\ln \mx / M_Z \right)^n$ are automatically included, where $\alpha$ is a
generic gauge or Yukawa coupling. It has been recognized quite early
\cite{10,49} that the radiative corrections described by the RGE can induce
spontaneous breaking of the $SU(2) \times U(1)_Y$ gauge symmetry by driving
some combination of the squared Higgs mass parameters $\mu^2_i$ of
eq.(\ref{e18}) to negative values. This is due to the effect of the Yukawa
couplings, which tend to reduce the squared masses of scalar fields. The Yukawa
sector therefore plays a crucial role in these models.

In a recent paper \cite{DN1} we studied radiative gauge symmetry breaking and
the particle spectrum in minimal SUGRA in some detail, taking care to
incorporate the effects of the Yukawa couplings of the $b$ quark and $\tau$
lepton, which can be quite important if $|\tanb| \gg 1$. We later
showed \cite{50} how to incorporate the ``finite'' (i.e., without
$\ln \mx / M_Z$ enhancement) radiative corrections to the Higgs sector in this
scheme. In particular we demonstrated that for most purposes these
radiative corrections can be made negligibly small by a proper choice of
the scale $Q_0$ where the RG running is terminated; the only exception is the
mass $m_h$ of the light Higgs scalar where the corrections have to be
included explicitly. (A similar result had been obtained previously in
ref.\cite{51}.) For more details we refer the reader to these papers, as well
as to earlier work on this subject \cite{10,49}; here we only give a brief
summary of the relevant properties of the spectrum.

As stated above, the squared mass of all sfermions at the GUT scale is
simply given by $m^2$. In case of the superpartners of the first two
generations
the  only sizable radiative corrections to the masses involve the gauge
interactions.  The breakdown on $SU(2) \times U(1)_Y$ gauge symmetry also
has some effect on sfermion masses. All these contributions can quite easily
be computed analytically; one has \be \label{e19}
m^2_{\tilde f_i} = m^2 + d_i M^2 + \cos \! 2 \beta
(I_{3,i} - Q_i \sin^2 \! \theta_W) M_Z^2,
\ee
where $I_{3,i}$ and $Q_i$ are the third component of the weak isospin and
electric charge of the sfermion $\tilde f_i$, respectively. The positive
constant $d_i$ is determined by the gauge quantum numbers of the sfermion;
numerically, it is about 6 for squarks, 0.5 for $SU(2)$ doublet sleptons and
0.15 for $SU(2)$ singlet sleptons with hypercharge 1.

The gaugino masses $M_i$ are all equal to $M$ at scale \mx; their $Q$
dependence is identical to that of the gauge couplings $\alpha_i$: \be
\label{e20}
M_i(Q) = \frac {\alpha_i(Q)} {\alpha_i{M_X}} M, \ee
which immediately implies eq.(\ref{e5}). Numerically, $M_3 \simeq 3 M, \
M_2 \simeq 0.84 M$ and $M_1 \simeq 0.43 M$.\footnote{In our numerical
calculations we use proper on--shell masses, i.e. $m_{\tilde f_i} \equiv
m_{\tilde f_i}(Q = m_{\tilde f_i})$ and similar for gluinos. The coefficients
in eqs.(\ref{e19}) and (\ref{e20}) then depend on the masses themselves,
rather than being simple constants. This can change the masses of strongly
interacting sparticles by as much as 20\%; e.g. for 1 TeV gluinos,
$m_{\tilde g} \simeq 2.5 |M|$, rather than $3 |M|$.}

Eqs.(\ref{e19}) and (\ref{e20}) have been taken into account in some previous
analyses \cite{20,21,19a,23,24} of LSP relic densities. Unfortunately the
effects of the Yukawa couplings are not so easily treated analytically.
These effects are important for the masses of third generation sfermions as
well as Higgs bosons. Indeed, Yukawa couplings affect sfermion masses already
at tree level \cite{45}; they lead to mixing between $SU(2)$ doublet and
singlet
sfermions, and give rise to additional supersymmetric diagonal mass terms.
These
effects are especially important for stop squarks, which obviously have the
largest Yukawa couplings; this has been included in the analysis of
ref.\cite{22}. However, for $|\tanb| \gg 1$, sbottom and especially stau
mixing also becomes important \cite{DN1}. Moreover, the Yukawa couplings reduce
the nonsupersymmetric diagonal mass terms from the values predicted by
eq.(\ref{e19}). The net effect is that the lighter stop and stau eigenstates
can be substantially lighter than the other squarks and sleptons,
respectively; indeed, no strict lower bound on these masses could be
given even if the masses of first generation sfermions were known. The
maximal reduction of the light sbottom mass is not quite as large, but can
still amount to 20 -- 30 \% if $|\tanb| \simeq m_t/m_b$. Since tree--level
and loop effects tend to cancel for the heavier eigenstates of third generation
sfermions, their masses are usually not very different from those of the
corresponding sfermions of the first two generations.

The Yukawa couplings also affect the Higgs masses via the RGE; as explained
above, this effect is at the heart of the radiative gauge symmetry breaking
mechanism. First of all, it should be noted that the masses of the heavier
Higgs bosons are related to the sfermion masses via eq.(\ref{e18}); this
equation only holds at scale \mx, but it shows that in SUGRA models one
cannot treat Higgs boson and sfermion masses as independent free parameters.
As well known \cite{52}, the masses of the $P, \ H$ and $H^+$ states are
essentially determined by $m_P$, which is simply given by \be \label{e21}
m_P^2 = \mu_1^2(Q_0) + \mu_2^2(Q_0); \ee
as discussed above, eq.(\ref{e21}) also holds to good approximation
after inclusion of 1--loop radiative corrections if the scale $Q_0$ where the
RG running is terminated is chosen properly, i.e. $Q_0 \simeq m_{\tilde q}$.
Notice that only the Higgs doublet $H_2$ couples to top quarks; moreover,
the requirement \be \label{e22}
\langle H_1^0 \rangle^2 + \langle H_2^0 \rangle^2 = 2 M_W^2/ g^2 \ee
immediately implies \cite{10} $\mu_2^2(Q_0) > -M_Z^2/2$. If the $b$ and
$\tau$ Yukawa couplings are neglected, the nonsupersymmetric contribution to
$\mu_1^2$ runs just like an $SU(2)$ doublet slepton mass, i.e. it
{\em increases} when going from $Q = \mx$ to $Q = Q_0$; in addition the
positive supersymmetric contribution $\mu^2$ has to be added \footnote{Note
that this is the square of the real parameter $\mu$, which renormalizes
multiplicatively; $\mu^2$ is therefore indeed always positive.}. This
argument shows that $m_P$ can only be smaller than the slepton masses if the
$b$ and $\tau$ Yukawa couplings are sizable, i.e. if $|\tanb| \gg 1$.
The exact numerical expression for $m_P$ can be approximated by \cite{DN1}
\be \label{e23}
m_P^2 = \frac {M_Z^2}{2} \left( \cot \! \beta - 1 \right) + \left( m^2 +
0.52 M^2 + \mu^2(Q_0) \right) \left( \frac {1} {\sin^2 \! \beta} - \frac
{a^2} {\cos^2 \! \beta} \right), \ee
where $a \simeq 1/45 - 1/35$ depends on the ratios $M/m, \ A/m$ as well as on
the top mass $m_t$.\footnote{In ref.\cite{DN1} we gave a somewhat larger
numerical value of $a$, because we under--estimated the effect of the
running of the $b$ quark mass between $Q_0$ and $m_b$. Since a smaller
$m_b(Q_0)$ also implies a larger upper bound on $|\tanb|$, our results for
sparticle masses, including $m_{\tilde b}$, remain valid if one rescales
$\tanb \to 1.15 \ \tanb$ in the region $|\tanb| \gg 1$.}

The mass of the lightest Higgs scalar $h$ in general depends on all parameters
of the model in a complicated way. However, in the important limit $m_P^2 \gg
M_Z^2$ the situation simplifies greatly, and one finds \cite{50}: \be
\label{e24}
m_h^2 = M_Z^2 \cos^2 2 \beta + \Delta_{22} \sin^2 \! \beta + {\cal O}
\left( \frac {M_Z^2}{m_P^2}, \frac{m_b^2}{m_t^2} \right). \ee
Here $\Delta_{22}$ describes the leading radiative corrections from
top--stop loops. It grows like the fourth power of $m_t$, but depends only
rather mildly on the values of the SUSY breaking parameters; in the limit
$m_{\tilde q}^2 \gg m_t^2$ it grows $\propto \ln m_{\tilde q}/ {m_t}$.

The SUGRA--imposed constraint that is most difficult to treat analytically
follows from the almost obvious observation that in the radiative gauge
symmetry breaking scenario the vevs of the Higgs fields can be computed from
the input parameters at the GUT scale, i.e. $\langle H_{1,2}^0 \rangle$
are functions of $m, \ M, \ A, \ \mu$ and the set of Yukawa couplings $h_t, \
h_b, \ h_{\tau}$. The condition (\ref{e22}) therefore leads to a relation
between these parameters. This relation cannot be expressed in closed
form once the effects from $h_{b, \tau}$ are included, but it can
approximately be written as \cite{DN1} \be \label{e25}
\mu^2(Q_0) \simeq \frac { \tan^2 \! \beta + 1} { \tan^2 \! \beta - 1} X_2
- m^2 - 0.52 M^2 - M_Z^2/2, \ee
where $X_2$ describes the effect of the RG--running due to the top quark
Yukawa coupling. An approximate expression for $X_2$ is\footnote{In
ref.\cite{DN1} we gave a somewhat larger expression for $X_2$, since we
had chosen a rather small value for $Q_0$, i.e. $Q_0 = M_Z$; eq.(\ref{e26})
is valid for $Q_0 \simeq 300$ GeV.}
\be \label{e26}
X_2 \simeq \left( \frac {m_t} {150 \ {\rm GeV}} \right)^2
 \left\{ 0.9 m^2 + 2.7 M^2 + \left[ 1 -
\left( \frac {m_t} {190 \ {\rm GeV}} \right)^3 \right]
\left( 0.24 A^2 + M A \right) \right\} \ee
Eq.(\ref{e25}) usually works to 10\% accuracy, but eq.(\ref{e26}) might
deviate by as much as 20\% from the exact numerical result; nevertheless these
expressions are quite useful to gain some insight into the relation between
the input parameters.

Strictly speaking, eq.(\ref{e25}) does not provide the searched--for relation
between input parameters at scale \mx\ on the one hand and $M_Z$ on the other,
since it still depends on \tanb, which is itself a complicated function of the
input parameters. (In fact, this is where the dominant effect from the other
Yukawa couplings enters.) However, $\mu^2$ becomes essentially independent of
\tanb\ if $\tan^2 \! \beta \gg 1$, in practice for $|\tanb| > 3$ or so.
Another complication arises because a given set of $m, \ M, \ A$ and $m_t$
often allows up to three different solutions \cite{DN1} of the equation that
determines \tanb\ (and hence the Yukawa couplings); these solutions differ in
both sign and magnitude of \tanb.\footnote{In this scheme the sign of \tanb\ is
is determined dynamically via the RGE. On the other hand, the couplings listed
in ref.\cite{7} and \cite{38} have been derived under the assumption $\tanb >
0$. Fortunately the masses and mixings of the neutralino and chargino
eigenstates only depend on the sign of the product $M \cdot \mu \cdot \tanb$; a
sign in \tanb\ can therefore always be ``rotated'' into $M$ or $\mu$.} We
therefore present results always for fixed $\tanb, \ M$ and $m$, rather than
fixing the values of the SUSY breaking parameters. Since consistent solutions
only exist for $h_t > h_b$, the allowed range of \tanb\ in this model is
restricted to $1 < |\tanb| < m_t/m_b$.

Unfortunately another two--fold ambiguity occurs when $\mu$ and $A$
are adjusted such that $M_Z$ and \tanb\ have their desired values. In one of
these solutions $\mu$ is of the order of $m$ and $M$. However, unless $m_t$ is
very large there is also a second solution with $\mu^2 \ll m^2, \ M^2$; indeed,
this is the ``small $\mu$'' solution which has been discussed in the first
analyses \cite{10} of radiative gauge symmetry breaking. However, by now this
kind of solution is quite severely constrained. First of all, we know from SUSY
searches at LEP \cite{41} that $|\mu(Q_0)| > 40$ GeV; since these solutions
typically have $|\mu|$ much smaller than $m$ and $M$, this constraint implies
that most sparticles must be quite heavy for these solutions to be acceptable.
Moreover, for $m_t > 155$ GeV the small$-\mu$ solutions disappear altogether,
since then the effect of the top Yukawa coupling always drives $\mu^2_2(Q_0)$
below $-M_Z^2/2$ unless it receives a sizable, positive contribution
$+ \mu^2$, see eq.(\ref{e18}).

At this point some comments on fine tuning might be appropriate. It should be
quite obvious that the ``natural'' scale for the vevs of the Higgs fields is
set by the dimensionful parameters of the Higgs potential, which in turn are
roughly of the order of typical sparticle masses, as can be seen from
eqs.(\ref{e21}) and (\ref{e25}). Therefore some fine tuning of parameters will
be necessary \cite{31,16} to achieve $M_Z^2 \ll m^2 + M^2$. This provides a
strong argument that sparticles should not be much heavier than ${\cal O}$(1
TeV), but this argument cannot be translated into strict upper bounds on
sparticle masses. Since one of the motivations of this study is to see whether
cosmology might provide us with such bounds, we do not impose any ``a priori''
upper bounds on the SUSY breaking parameters in our analysis. One might also
argue that large ratios of the mass parameters of the model are ``unnatural'',
but the Yukawa sector shows that large ratios of ``fundamental'' parameters can
indeed occur. Of course, one ultimately hopes to understand the origin of the
dimensionful parameters $m, \ M, \ A$ and $\mu$ better, e.g. in the framework
of superstring theories \cite{53}. However, at present it seems safer to pursue
an ``agnostic'' approach, and try to cover the {\em entire} experimentally
allowed parameter space.

This allowed region is defined via the following experimental constraints:
\ben \label{e27} \beq
m_{\tilde {e}_{R,L}, \tilde{\tau}_1, \tilde{t}_1, \tilde{\chi^+}} &> 45 \
{\rm GeV}; \label{e27a} \\
m_{\tilde{\nu}} &> 40 \ {\rm GeV}; \label{e27b} \\
\sum_{i,j=1}^4 Br(Z \to \tilde{\chi}_i^0 \tilde{\chi}_j^0) &< 5 \cdot 10^{-5};
\label{e27c} \\
\Gamma(Z \to \tilde{\chi}_1^0 \tilde{\chi}_1^0) &< 12 \ {\rm MeV};
\label{e27d} \\
m_{\tilde g} &> 120 \ {\rm GeV}; \label{e27e} \\
m_{\tilde{\tau}_1} &\geq \mchi. \label{e27f} \eeq \een
The bounds (\ref{e27a})--(\ref{e27d}) directly follow from LEP limits on
sparticle production \cite{41} as well as on the invisible width of the $Z$
boson; here $\tilde{\chi}^+$ and $\tilde{\chi}_i^0$ stand for a generic
chargino and neutralino state, with $\tilde{\chi}_1^0 \equiv \chi$. The bound
(\ref{e27e}) is a conservative interpretation of the preliminary CDF search
limits \cite{42} after inclusion of cascade decays \cite{43}. Finally,
(\ref{e27f}) follows directly from the requirement that the LSP should not be
charged, as discussed in the introduction. Further experimental constraints
follow from the unsuccessful search of Higgs bosons; we have incorporated a
parametrization of the ALEPH bound \cite{higb} in our list of conditions.

In addition to imposing these
experimental constraints, we also discard combinations of parameters that lead
to deeper lying minima of the scalar potential that break charge and/or color;
this requirement excludes combinations with $A^2/ \left( m^2 + M^2 \right) \gg
1$ \cite{falvac}. Finally, we demand that the scalar potential should be
bounded from below at scale $Q = M_X$, which implies $\mu_1^2 + \mu_2^2 \geq
2 |\mu_3^2|$; eq.(\ref{e18'}) shows that this excludes the region
\be \label{e27'}
\frac {|B|}{2} - \sqrt{ \frac {B^2}{4} -1 } < |\mu(M_X)| <
\frac {|B|}{2} + \sqrt{ \frac {B^2}{4} -1}, \ee
if $|B| \equiv |A-m| \geq 2 m$. This constraint is effective at small
$|\tanb|$,
which implies large $\mu$, see eq.(\ref{e25}), and thus large and positive
squared Higgs mass parameters (\ref{e19}). One then needs large $|A/m|$ to
achieve spontaneous gauge symmetry breaking, since this accelerates the RG
running of the mass parameters; however, large $|B/m|$ also imply a large
excluded region (\ref{e27'}).
\setcounter{footnote}{0}
\section*{3. Examples}
We are now in a position to present some numerical results. The discussion of
sec. 2c showed that the model has 4 free parameters, which we chose to be
$m, \ M, \ m_t$ and \tanb; $\mu$ and $A$ are then fixed by the equations
describing the minimization of the Higgs potential, up to a possible discrete
ambiguity. Without loss of generality $m_t$ and $m$ can be chosen to be
positive, but $M$ and \tanb\ can have either sign.

Figs. 1 a--d show a first partial exploration of the parameter space of the
model. In these figures we have fixed $m=300$ GeV, which leads to
cosmologically interesting DM densities for a wide range of the remaining
parameters. In addition, in each figure we have kept $m_t$ and \tanb\ fixed,
and varied $M$. We find that for the chosen values of parameters only one
experimentally allowed solution for $A$ and $\mu$ exists. Moreover, since in
all these cases the mass of the LSP increases monotonically with $|M|$,
we present our results as a function of \mchi; this simplifies the
identification of the various $s$--channel poles and of the thresholds
where new annihilation channels open up. The starting point of all curves
in figs. 1 is determined by the LEP constraints (\ref{e27a}), (\ref{e27c})
and (\ref{e27d}).

We see that in the limit of large \mchi\ all curves become almost identical.
The reason is that in this case we always have $|\mu| > |M_1| \gg M_Z$, so
that the LSP is bino--like. The dominant annihilation channel is then
$\chi \chi \to l^+ l^-$ via $\tilde{l}$ exchange, where $l$ is any lepton.
Eq.(\ref{e9}) shows that
squark exchange is suppressed \cite{21} by their large mass (\ref{e19}), and
the $Z$ and Higgs exchange contributions are suppressed by small couplings.
As discussed in sec. 2b, the production of
longitudinal gauge bosons $V_L$ and of the light Higgs boson $h$ are not
suppressed by powers of $M_Z/\mchi$, but are suppressed by powers of
$|M_1/\mu| \simeq 1/4 - 1/2$ for the examples of figs. 1; moreover, all
neutralino couplings relevant for these final states originate (either
directly or via the equivalence theorem) from the Higgs--higgsino--gaugino
$U(1)_Y$ gauge interaction, which involve fields with hypercharge $|Y| = 1/2$,
while the $SU(2)$ singlet leptons have $Y=1$. The gauge and Higgs boson
final states therefore only contribute a few \% in the region $\mchi > 200$
GeV ($> 300$ GeV for the dashed curve in fig. 1c; see below).

The behaviour of the curves in this region can therefore be understood
semi--quantitatively from the $\tilde{l}_R$ exchange contribution alone:
\be \label{en1}
\sigma_{\rm ann} \propto \frac {m^2_{\chi}}
{\left( m^2_{\tilde{l}_R} + m^2_{\chi} \right)^2} \left[
\left( 1 - \frac {m^2_{\chi}} {m^2_{\tilde{l}_R} + m^2_{\chi}} \right)^2
+ \frac {m_{\chi}^4} {(m^2_{\tilde{l}_R} + m^2_{\chi})^2} \right], \ee
where the parenthesis results from the Taylor expansion of the propagator in
powers of the velocity $v$. SUGRA predicts $m^2_{\tilde{l}_R} \simeq m^2 + 0.83
m^2_{\chi}$ for the given case of a bino--like LSP, see eqs.(\ref{e19}) and
(\ref{e20}); eq.(\ref{en1}) then leads to a maximum of the annihilation cross
section, i.e. a minimum of the relic density, at $\mchi \simeq 0.6 m \simeq
180$ GeV for the parameters of figs. 1. This maximum is very broad; the
$\tilde{l}_R$ exchange contribution falls to 50\% of its maximal value at
$\mchi \simeq 1.52 m$, just beyond the end of the region shown in figs. 1.
Indeed, in Figs. 1a,b, \oh\ at \mchi\ = 430 GeV is about twice as large as at
the minimum. Eq.(\ref{en1}) also predicts the annihilation cross section to
fall at small values of \mchi, dropping to half the maximum value at $\mchi
\simeq 0.25 m$. However, in many cases our assumption that $\tilde{l}_R$
exchange dominates the total annihilation cross section is no longer valid in
this region.

Going towards smaller values of \mchi, the first prominent structure one
encounters is the \ttbar\ threshold. It is most prominent for the dashed curve
in fig. 1c, since this combination of parameters leads to the smallest value of
$|\mu|$; for a given value of $M$, a smaller $|\mu|$ means larger higgsino
admixtures to the LSP, see eq.(\ref{e6a}), and hence larger couplings to $h$
and $Z$ bosons. For most of cases shown in figs. 1 the contribution of the
\ttbar\ final state is suppressed by destructive interference between
$\tilde{t}$ and $Z$ exchange contributions, which have approximately equal
magnitude but opposite signs here. On the other hand, top production is not
$p-$wave suppressed unless $m^2_{\chi} \gg m_t^2$, unlike the production of
massless fermions. The importance of the \ttbar\ final state compared to light
fermions is therefore enhanced by a relative factor $x_F/3 \simeq 10$, see
eq.(\ref{e3}).

In the region below the \ttbar\ threshold the differences between the
various curves start to become more pronounced. Many of these differences
can be understood from eqs.(\ref{e25}) and (\ref{e26}), which show that
increasing $m_t$ and decreasing \tanb\ both imply larger values of $\mu$,
which leads to smaller couplings of the LSP to Higgs and gauge bosons, and
suppresses contributions from the $t-$channel exchange of higgsino--like,
heavier neutralinos. This explains why the $WW$ and $ZZ$ thresholds, as well
as the minima at $\mchi = M_Z/2$ and at $\mchi = m_h/2$, are more prominent
in fig. 1a than in 1b, and why \oh\ at small \mchi, where $Z$ exchange
diagrams dominate the annihilation cross section, is considerably smaller
in figs. 1c,d than in 1a,b. Moreover, $m_h$ increases with increasing
$m_t$ and increasing $|\tanb|$, so that the position of the $h-$pole tends
to move to larger values of \mchi\ as we go from fig. 1a to 1d.

The depth and width of the minimum at $\mchi = m_h/2$ depends quite
sensitively on the choice of parameters. Eq.(\ref{e23}) shows that in figs. 1
we always have $m_P^2 \gg M_Z^2$; in this limit the $h \chi \chi$ coupling
\cite{38} becomes for a bino--like LSP: \be \label{e28}
g_{h \chi \chi} = \frac {g'}{2} \frac{ M_Z \sin \! \theta_W (M_1 + \mu
\sin \! 2 \beta)} {M_1^2 - \mu^2} +{\cal O}\left( \frac {M_Z^2} {M_1^2 -
\mu^2}, \frac {M_Z^2} {m_P^2} \right). \ee
Notice that the two terms in the numerator tend to cancel if $\pro < 0$;
this explains why the minimum at $\mchi = m_h/2$ is narrower and shallower
for the dashed curves in figs. 1 than for the solid ones. However, we
remind the reader that estimating \oh\ from eqs.(\ref{e1})--(\ref{e3}) can
lead to large errors in the vicinity of a very narrow pole. A more
careful treatment \cite{33} would lead to shallower minima, which are
broadened in the region below $m_h/2$; if $m_h > M_Z$, the relative
maximum between the two minima should therefore also be somewhat lower than
indicated in figs. 1. On the other hand, the $h-$ pole clearly affects only
a very limited region of parameter space; our overall conclusions do therefore
not depend on an accurate treatment of this pole.

Finally, the strength of the $Z \chi \chi$ coupling also depends quite
sensitively on the choice of parameters, including their signs. In this case
the ordering of the curves for $M>0$ and $M<0$ even depends on \tanb. For
\tanb\ = 2, a cancellation occurs \cite{39,40} in the neutralino (and chargino)
mass matrix if $\pro > 0$; the LSP then has substantial higgsino components if
$\mchi \leq M_Z/2$. In contrast, for small \tanb\ and $\pro < 0$ the LSP
remains dominantly a gaugino even if \mchi\ is very small. On the other hand,
for $M<0, \ \tanb = 15, \ \mu(Q_0)$ is quite small due to a strong cancellation
in the r.h.s. of eq.(\ref{e25}), especially for $m_t = 130$ GeV, fig. 1c. At
larger $|M|$, i.e. larger \mchi, this cancellation is less complete, but for
this choice of parameters $\chi$ remains dominantly a higgsino for $\mchi \leq
60$ GeV, and reaches 90\% bino--content only for $\mchi \geq 120$ GeV; for
$M>0$ and the same values of $m_t$ and \tanb, the LSP has already 97\%
bino--content at this mass. This explains why the two curves in fig. 1c differ
quite strongly even at rather large values of \mchi.

The strong dependence of some of the annihilation cross sections on the model
parameters is further illustrated by fig. 2, which shows a blow--up (on a
linear scale) of the $VV$ and $Zh$ threshold region for two of the curves
of figs. 1. Both curves show small shoulders at the $WW$ and $ZZ$ thresholds.
As explained above, these thresholds are somewhat less pronounced for the
case $m_t = 160$ GeV, due to the larger value of $\mu$. However, since for
fixed
\mchi,
$\mu$ only increases by approximately 25\% as $m_t$ is increased from 130 to
160 GeV, this effect is not very large; see also eq.(\ref{e12}). This rather
small change of parameters suffices, however, to reduce the $\chi \chi \to hh$
cross section by as much as a factor of 6! For $m_t = 130$ GeV, the exchange
of the lighter, gaugino--like and heavier, higgsino--like neutralinos
gives contributions of approximately equal size and equal sign to the matrix
element. In other words, the terms $\propto d_8$ and $\propto d_9$ in
eq.(\ref{e16}) have about equal magnitude here; notice that the larger $SU(2)$
gauge coupling can only enter via $d_9$ in case of a bino--like LSP. Going
to larger $m_t$ does not only increase $\mu$ for fixed $M$, it also increases
$m_h$ via the radiative correction $\Delta_{22}$ of eq.(\ref{e24}); this
necessitates an increase of $M$, and thus a further increase of $\mu$, in
order to reach the $hh$ threshold. The contributions $\propto d_9$ are
therefore almost negligible for the dashed curve in fig.2. Indeed, the
decrease of this curve after the maximum at $\mchi = 62$ GeV has nothing to do
with any thresholds; rather it is caused by the increase of the slepton
exchange contribution to the $l^+l^-$ final state, see eq.(\ref{en1}).
Finally, we remind the reader that a cancellation occurs in
the contribution to the $hh$ final state from higgsino exchange if $\pro < 0$,
see eq.(\ref{e17}); this contribution is also suppressed for large \tanb\ if
$|\mu| \gg |M_1|$. The $hh$ threshold is therefore all but invisible in most
curves in figs. 1.

It has been noted in ref.\cite{33} that our estimate of \oh\,
eqs.(\ref{e1})--(\ref{e3}), becomes unreliable in the vicinity of a threshold
for a final state which quickly dominates the total annihilation cross section.
The example given there was exactly the $hh$ final state, which for certain
combinations of parameters can dominate the total cross section by a large
factor. However, we find that such a situation never occurs for bino--like
or mixed LSP's in minimal SUGRA; in this case the $hh$ threshold, as well as
all other thresholds, is never much more pronounced than for the solid
curve in fig. 2, and usually the thresholds are much less important, as can
be seen from figs. 1. We do therefore not expect the error introduced by our
approximate treatment to exceed 10\% just below threshold, and it
should be much smaller everywhere else.

So far we have concentrated on examples where the LSP is dominantly a bino,
the exception being the dashed curve in fig. 1c. Since the LSP will only
be higgsino--like if $|\mu| < |M_1| \simeq 0.43 |M|$, one obviously needs
quite large values of $|M|$ to get a higgsino--like LSP with mass
substantially above $M_Z$. The SUGRA constraints then imply that one also
needs large $m$, since for $M^2 \gg m^2$ eqs.(\ref{e25}) and (\ref{e26})
always yield $|\mu| \geq |M_1|$ for experimentally allowed values of $m_t$.
In fig. 3 we have therefore chosen $m=2$ TeV, and present results
for two different
choices of $M$ and $m_t$. In this figure, $\mu, \ A$ and \tanb\ all vary
along the $x-$axis, with $\mchi \simeq |\mu|$. As in figs. 1, we see that
\oh\ is essentially independent of most parameters if the LSP is a heavy,
almost pure state, here an almost pure higgsino. Of course, once $|\mu| >
|M_1|$ the LSP will become bino--like again, and will thus have a very small
annihilation cross section, since the very large $m$ implies very heavy
sfermions; this explains the steep rise of the dashed curve at $\mchi \simeq
430$ GeV.

Notice that this curve does not have a relative minimum in the
region $|\mu| \simeq |M_1|$, even though here both the higgsino and
gaugino components of the LSP are large; this seems to be in conflict
with results of refs.\cite{18,24a}. However, in our case eq.(\ref{e23})
implies that all Higgs bosons except $h$ are very heavy, so that most
final states containing Higgs bosons are not accessible; moreover,
contributions
from $P-$exchange in the $s-$channel are suppressed, even though the $P \chi
\chi$ coupling is large for a mixed LSP. The $\chi \chi \to \ffbar$ cross
section does show a maximum in the region where $\chi$ is a mixed state, due
to the contributions from $Z$ and $h$ exchange, but this is not sufficient
to compensate the rapid decrease of the $VV$ and $Zh$ cross sections.

For $\mchi < M_W$ the annihilation of higgsino--like LSP's is dominated by
$Z-$exchange diagrams. Eqs.(\ref{e9'}) and (\ref{e7}) show that
$g_{Z \chi \chi} \propto M^2_Z / (\mu M_1)$ in this case; the $\chi \chi$
annihilation cross section is therefore about 4 times smaller for $M=2$ TeV
than for $M=-1$ TeV. However, the results of fig. 3 are quite misleading in
this region. We had already mentioned above that eqs.(\ref{e1})--(\ref{e3})
are not valid \cite{33} close to a threshold where a new channel opens up
which quickly dominates the total annihilation cross section, as is the
case for the $WW$ threshold here; this is because we ignored the possibility
that LSP's with mass below $M_W$ can annihilate into $WW$ pairs if they have
sufficient kinetic energy. Therefore we have overestimated the relic density
in the region just below and at the $WW$ threshold.

Moreover, as already discussed in sec. 2b, a light higgsino--like LSP always
implies the existence of a second higgsino--like state $\chi'$ whose mass is
quite close to that of the LSP. Since the $Z \chi \chi'$ coupling is {\em not}
suppressed, unlike the $Z \chi \chi$ coupling, the $\chi \chi'$
co-annihilation cross section is much larger than $\sig(\chi \chi)$, see
eq.(\ref{e11}). Furthermore, $\chi \chi'$ co-annihilation is {\em not}
$p-$wave suppressed even if the final state fermions are massless. Using
the formalism of ref.\cite{33}, we estimate that inclusion of $\chi \chi'$
co-annihilation would reduce \oh\ by a factor \be \label{e29}
K \simeq \left( 1 + \frac{x_F}{3} \frac {2} {\epsilon^2}
e^{-x_F \Delta} \right) / \left( 1 + e^{-x_F \Delta} \right)^2. \ee
Here $\Delta \equiv |(|m_3|-|m_4|)/\mu|$ is given by eq.(\ref{e8}) and
$\epsilon$ by eq.(\ref{e7}). The exponential factor describes the Boltzmann
suppression of the $\chi'$ density at freeze--out; we find $x_F \simeq 25$ in
this case. The factor of $x_F/3$ in front of the second term in the numerator
of eq.(\ref{e29}) has been included to estimate the $s-$wave enhancement (or,
more accurately, lack of $p-$wave suppression) of the co-annihilation process,
and the 2 is a statistics factor \cite{33}. Numerically we find $K \simeq 4$
(80) for $|M| = 1$ TeV and $|\mu| = 50$ (80) GeV; for $|M|=2$ TeV the
corresponding numbers are 230 and 1500, respectively. Of course, these
estimates could easily be off by a factor of 2 or so; nevertheless, taken
together with sub--threshold $\chi \chi$ annihilation into $W$ pairs, these
large suppression factors allow us to conclude that the relic density of
higgsino--like LSP's will always be uninterestingly small unless $\mchi \geq
500$ GeV or so.\footnote{Of course, co--annihilation will also occur for $\mchi
> M_W$; indeed, the relative mass splitting $\Delta$ between the higgsino
states, and hence the Boltzmann suppression of the co-annihilation
contribution, will be (much) smaller than for light higgsinos. On the other
hand, $\sig (\chi \chi')$ should not be much bigger than $\sig (\chi \chi)$ in
this case, since $\sig (\chi \chi)$ is dominated by annihilation into a pair of
gauge bosons, which occurs with full gauge strength; in the region $\mchi >
M_W$ co--annihilation should therefore not change the result of fig. 3 much.}

We had seen in figs. 1 that over a wide region of parameter space the relic
density of a heavy, bino--like LSP depends only very little on $m_t$, \tanb\
and the sign of $M$. However, as already pointed out in ref.\cite{DN1}, this
is no longer true for very large values of $|\tanb|$. This is illustrated in
fig. 4, where we show \oh\ as a function of \tanb\ for fixed $m, \ M$ and
$m_t$; the parameters are chosen such that $\chi$ is bino--like. The full
line shows the SUGRA prediction including the contributions from the $b$ and
$\tau$ Yukawa couplings to the neutralino--fermion--sfermion interactions
as well as to the RGE. These latter contributions reduce $m_P$, as
shown in eq.(\ref{e23}); they also reduce the masses of the lighter $\tilde{b}$
and $\tilde{\tau}$ eigenstates by mixing between $SU(2)$ singlet and doublet
states, and by reducing the diagonal entries of their mass matrices.

Fig. 4 shows that both these effects are quite important. When \tanb\ is
increased sufficiently, one eventually has $m_P = 2 \mchi$; for the parameters
of fig. 4 this happens at $\tanb \simeq 35$. This results in
a very strong enhancement of the $\chi \chi \to b \overline{b}, \ \tau^+
\tau^-$
cross sections via the exchange of a pseudoscalar Higgs boson. Note that the
$P b \overline{b}$ and $P \tau^+ \tau^-$ couplings increase $\propto \tanb$,
so that the total decay width $\Gamma_P \propto \tan^2 \! \beta$. For
$\tanb = 35$ the mass to width ratio of $P$ is therefore similar to that of the
$Z$ boson, so that our estimate of \oh\ should be quite reliable even in the
pole region \cite{33}.

The long dashed curve has been obtained by artificially keeping $m_P$ constant
at the value SUGRA predicts for $\tanb = 2 \ (\simeq 780$ GeV); this also
implies that $m_{H^+}$ and $m_H$ are kept (approximately) constant. However,
the
$\tilde{b}$ and $\tilde{\tau}$ masses are still allowed to vary with \tanb\ as
predicted by SUGRA, and the Yukawa contributions to the neutralino couplings
are included; we see that this suffices to reduce \oh\ by approximately a
factor of 3 at the largest allowed value of \tanb. The reduction of
$m_{\tilde{\tau}_1}$ is the dominant effect here, since for the given choice of
$m$ and $M$ the $\tilde{b}$ squarks are almost twice as heavy as the
$\tilde{\tau}$ sleptons, and have smaller hypercharge. Because for the given
choice of parameters one has $m^2_{\chi} \ll m^2_{\tilde f}$, the annihilation
cross section is essentially proportional to $\sum_i Y_i^4 / m^4_{\tilde
{l}_i}$ here, where $i$ is a generation index; at small \tanb, all three
generations contribute almost equally, but for large \tanb\ the sum is
dominated by the contribution from the 3rd generation. The reduction of \oh\ by
a factor of 3 then corresponds to a reduction of $m_{\tilde{\tau}_1}$ by only a
factor of 1.63.  An even larger reduction of the $\tilde{\tau}$ mass is not
possible here since a further increase of \tanb\ would lead to $m^2_P < 0$; as
shown in ref.\cite{DN1}, this constraint implies $|\tanb| \leq m_t(Q_0) /
m_b(Q_0)$. Finally, the short dashed curve has been obtained by ignoring all
effects from the $b$ and $\tau$ Yukawa couplings; it is a few \% above the
other curves even at small \tanb\ since the effects of sfermion mixing are not
entirely negligible even here. We see that in this case, which approximates the
usual analyses based on a global SUSY model with independent sparticle masses
at the weak scale, \oh\ does indeed only depend very little on \tanb.

Using the insight gained from figs. 1--4 it is now quite straightforward to
interpret the contour plots of figs. 5 and 6. Each of these figures is for
fixed values of $m$ and $m_t$ and for a given choice of the signs of $M$ and
\tanb. In figs. 5 a--d we choose $m=250$ GeV, $m_t$ = 140 GeV, and explore the
plane of $M$ and \tanb\ for all 4 combinations of signs. Solid and long dashed
lines are contours of constant \oh\ = 1 and 0.25, respectively. The short
dashed
curves in fig. 5a are lines of constant \oh\ = 0.025; since these contours
cluster very narrowly around the $Z$, $h$ and $P$ poles, in the other figures
we have merely indicated the position of these poles with the short dashed
lines.

Finally, the region outside the dotted curves is excluded by the experimental
and theoretical constraints discussed at the end of sec. 2c. In the region of
small $|M|$ the most important constraints are the LEP search limits
(\ref{e27a})--(\ref{e27d}) as well as the gluino mass bound (\ref{e27f});
for small $|\tanb|$ the LEP limit on the production of the light scalar Higgs
boson also plays a role. The lower bound on $|\tanb|$ is determined by the
Higgs search bound as well as the requirement that the Higgs potential should
be bounded from below even at the GUT scale, as discussed in sec. 2c below
eq.(\ref{e27'}). The region of large $|\tanb|$ and small and moderate values
of $|M|$ is limited by the LEP bounds on associate $hP$ production, which
is practially equivalent to requiring $m_P^2 > 0$ here, since the overall
mass scale ($m$) is chosen quite high in these figures. Finally, in the
region of large $|M|$ (or, more accurately, large $|M/m|$) the parameter
space is limited by the requirement of a neutral LSP, eq.(\ref{e27f}); this
bound is especially important for large $|\tanb|$, since
$m_{\tilde{\tau}_1}$ is smaller there, as discussed in connection with
fig. 4.

The gross features are the same in all 4 figures. At large $|M|$, \oh\ exceeds
1, as already shown in figs. 1; this also happens at small $|\tanb|$ and
small $|M|$, below the $Z$ and $h$ poles. Finally, there is a third region
where the $\chi$ relic density is unacceptably large, covering the region
between the $Z$ and $h$ poles and the $VV, \ Zh$ and $hh$ thresholds at small
and moderate values of $|\tanb|$. The extension of these last two excluded
regions does depend on the signs of $M$ and \tanb, however. We have already
seen
that the contribution from $h$ exchange  and the annihilation cross
section into the $hh$ final state are much smaller if $\pro < 0$, fig. 5b,c;
in this case the line \oh\ = 0.25 can even cross the $h$ pole. Moreover, for
$|M| \leq 300$ GeV negative values of \tanb\ (figs. 5b,d) require rather
large, positive values of $A$; if in addition $M>0, \ X_2$ of eq.(\ref{e26})
and hence $|\mu|$ become large, leading to small higgsino components of the LSP
and thus large values of \oh\ in fig. 5b. In fig. 5d we have $M<0$, however,
which results in a partial cancellation in $X_2$ and much smaller values of
$|\mu|$; together with sizable contributions from $h$ exchange and $hh$
production this explains the smallness of the cosmologically
excluded region for this choice of signs. On the other hand, small or
moderately
positive values of \tanb\ imply $A \simeq 0.5$ in the region $|M| \leq m$,
reducing the impact of this parameter on $X_2$ and $\mu$; furthermore,
choosing $M>0$ now results in larger contributions from the light Higgs boson.
The differences between figs. 5a and 5c in the experimentally allowed region
are therefore smaller than those between figs. 5b and 5d.

Because of the cancellation in the neutralino mass matrix discussed in
connection with figs. 1, LEP constraints from neutralino searches lead to a
more stringent limit on $|M|$ if $\pro > 0$; this explains why the
experimentally allowed, but cosmologically excluded region of small $|M|$ is
larger in figs. 5b,c than in 5a,d. Finally, as shown in ref.\cite{DN1}, one can
only achieve $|\tanb| \gg 1$ for sizable values of $\mu$ if $A>0$, and usually
$A>m$; choosing $M<0$ then reduces the effect of the Yukawa coupling on the
running of scalar masses, as exemplified by $X_2$, eq.(\ref{e26}). Somewhat
paradoxically, this reduces $m_P$, because in the relevant limit $\tan^2 \!
\beta \gg 1$ eqs.(\ref{e24}) and (\ref{e25}) imply $m_P^2 \propto X_2$. A
smaller $X_2$ also implies smaller $\mu$ and hence less stau mixing, which
again increases $m_{\tilde{\tau}_1}$. Altogether we thus see that choosing
$M<0$ leads to smaller values of $m_P$ and larger \mstau\ in the region of
large $|\tanb|$.  The region of parameter space to the right of the $P$ pole
that is allowed by the requirement $\mchi \leq m_{\tilde{\tau}_1}$ is therefore
somewhat larger for $M<0$, figs. 5b,d, than for $M>0$, 5a,c. One even finds
another small region with $\oh > 0.25$ at very large $|\tanb|$ and $M \simeq
-800$ GeV.

In figs. 6 we have chosen $M$ and \tanb\ to be positive, and study the effects
of varying $m_t$ (figs. 6a,b) or $m$ (6c,d). We have already seen that
increasing $m_t$ increases $|\mu|$, and thus also \oh\ if $\chi$ is bino--like.
Indeed we find larger cosmologically excluded regions in fig. 6a than in fig.
5a. Moreover, the fraction of the plane with $M < 200$ GeV where $\oh > 0.25$
is now much larger than before. This is partly due to the increase of the mass
of the light Higgs boson caused by the increase of $\Delta_{22}$ in
eq.(\ref{e24}). For $\tanb \geq 5$ the $h$ and $Z$ poles are now sufficiently
far apart to allow for a new region with cosmologically interesting DM density
in between these poles. Larger values of $|\mu|$ also imply \cite{DN1} more
$\tilde{\tau}$ mixing, which reduces $m_{\tilde{\tau}_1}$; at the same time
increasing $|\mu|$ implies larger values of $m_P$, see eq.(\ref{e23}). The
neutral LSP constraint (\ref{e27f}) therefore does no longer allow to choose
$|\tanb|$ so large that $m_P \simeq 2 \mchi$, except for a small stretch at $M
\simeq 130$ GeV; the effect of the reduction of $m_P$ and $m_{\tilde{\tau}_1}$
at large \tanb\ is nevertheless still quite pronounced in fig. 6a. Finally we
mention that the requirement that the top Yukawa coupling remains finite up to
scale \mx\ implies $|\tanb| \geq 2$ for $m_t = 170$ GeV.

Reducing $m_t$ from 140 to 110 GeV, fig. 6b, has obviously the opposite effect
as increasing it to 170 GeV. $|\mu|$ is now so small that the whole region of
small and moderate $|M|$ is cosmologically safe; however, this also implies
larger couplings of the lighter neutralino states to the $Z$ boson, so that the
LEP search limits rule out a much larger part of the plane than in fig. 5a.
(Recall that the signs of the parameters are such that cancellations occur in
the determinant of the neutralino mass matrix.) We already saw in fig. 2 that
the $hh$ contribution depends very sensitively on $M, \ \mu$ and \tanb; in the
given case it is large enough to create a small ``island'' with $\oh < 0.25$ at
$\tanb \simeq 2, \ M \simeq 200$ GeV. Moreover, for given values of $M$ and
\tanb, $m_{\tilde{\tau}_1}$ is now larger and $m_P$ is smaller than in fig. 5a,
so that a sizable region of parameter space to the right of the $P$ pole is
again allowed, including a substantial region where $\oh > 0.25$. The
relatively light pseudoscalar here even affects the cosmologically excluded
region at very large $M$, leading to a much steeper slope of the uppermost
solid line than in fig. 5a. Finally, the reduction of $|\mu|$ also implies that
the requirement (\ref{e27'}) of a bounded Higgs potential at scale \mx\ now
excludes a large region of parameter space at large $|M|$ and small \tanb.

In figs. 6c,d we have again chosen $m_t = 140$ GeV, but have varied $m$
compared to the value of fig. 5a. A larger $m$ means larger sfermion masses
and hence smaller contributions from sfermion exchange diagrams, which are
dominant at large $|M|$, as we have seen above. Indeed, we see that the
uppermost contour with \oh\ = 1, which (at least for small $|\tanb|$)
occured essentially in the same place
in the previous 6 figures, depends very strongly on $m$: For $m$ = 400
GeV the cosmologically safe region above the $VV, \ Zh$ and $hh$ thresholds
has disappeared completely; on the other hand, for $m$ = 125 GeV, fig. 6d,
all experimentally allowed combinations of $M$ and \tanb\ give $\oh < 1$, so
that the requirement that the relic LSP density should not overclose the
universe doesn't constrain the parameter space any further.
Recall that an increase of $m$ also implies an increase of
$|\mu|$ via eq.(\ref{e25}); this explains the differences in the region of
small $M$, below and around the $h$ pole, between figs. 5a, 6c and 6d, even
though sfermion exchange contributions are essentially negligible here.
Finally,
for $m=125$ GeV we observe a region with very small relic density, $\oh <
0.25$,
in approximately the same place that leads to $\oh > 1$ for $m=250$ GeV; this
once again demonstrates the strong dependence of the contributions from gauge
and Higgs final states on the parameters of the model.

This concludes our discussion of samples of the parameter space of the model.
We now attempt to derive bounds on sparticle masses or model parameters
from computations of the DM density.
\setcounter{footnote}{0}
\section*{4. Bounds}
We have seen in the previous section that very heavy LSPs tend to have small
annihilation cross sections. This is not surprising, since unitarity requires
the cross section to fall off at least like $1/m^2_{\chi}$ as $\mchi \to
\infty$, for fixed values of the other parameters. Indeed, we see from figs. 5
that for given $m$ and $m_t$ the requirement $\oh \leq 1$ imposes an upper
bound on $|M|$. Moreover, in the important case of a bino--like LSP away from
all ($Z$ and Higgs) poles, the resulting bound depends essentially only on $m$
and \mchi, as shown in figs. 1, 5 and 6. The SUGRA relations for sfermion
masses (\ref{e20}) imply that the sfermions with the largest hypercharge, the
superpartners $\tilde{l}_R$ of the right--handed leptons, also have the
smallest masses; they will therefore dominate the annihilation cross section
unless $M^2 \ll m^2$, as already pointed out in the previous section. From
eqs.(\ref{e20}) and (\ref{en1}) and the numerical result of figs. 1 that $\oh
\simeq 1$ for $m = 300$ GeV, \mchi = 180 GeV if $\chi$ is bino--like, one then
derives the approximate bound \be \label{e34}
\frac {\left( m^2 + 1.83 m^2_{\chi} \right)^2}
{m^2_{\chi} \left[ \left( 1- \frac {m^2_{\chi}} {m^2 + 1.83 m^2_{\chi}}
\right)^2 + \left( \frac {m^2_{\chi}} {m^2 + 1.83 m^2_{\chi}} \right)^2
\right] } \leq 1 \cdot 10^6 \ {\rm GeV}^2; \ee
this bound is only valid for a bino--like LSP away from poles. We already
mentioned in the previous section that for fixed $m$ the l.h.s. of (\ref{e34})
has a minimum at $\mchi = 0.6 m$. Plugging this into the bound (\ref{e34})
gives \be \label{e35}
m \leq 300 \ {\rm GeV} \ee
for any value of \mchi; this can also be read off figs. 1. Similarly, for given
\mchi\ the l.h.s. of (\ref{e34}) is minimized by choosing $m$ as small as
allowed by the constraint $\mlr \geq \mchi$, i.e. for $m^2 + 0.83 m^2_{\chi} =
m^2_{\chi}$. This immediately gives the bounds \ben \label{e36} \beq
\mchi &\leq 350 \ {\rm GeV}; \label{e36a} \\
|M| &\leq 825 \ {\rm GeV}, \label{e36b} \eeq \een
for any $m$. The bound (\ref{e36a}) is considerably stronger than the bound of
550 GeV given in ref.\cite{18}, because in that paper {\em all} sfermions were
allowed to have mass $m_{\tilde f} = \mchi$, which is not possible in minimal
SUGRA. In particular, a light stop greatly enhances the annihilation into
\ttbar\ pairs, which also makes a sizable contribution to the $s-$wave, ${\cal
O}(v^0)$ cross section.

We emphasize again that the bounds (\ref{e34}) -- (\ref{e36}) only hold for a
bino--like LSP away from poles. In particular, the bound (\ref{e35}) on $m$ can
be violated even for small $|\tanb|$ if \mchi\ is close to $M_Z/2$ or $m_h/2$,
see fig. 6c; in this case no useful bound on $m$ can be given. Moreover, figs.
5 and 6 show that for a given combination of $m, \ m_t$ and \tanb, the upper
bound on $|M|$ is weakest in the region close to the $P$ pole, if $m \geq 150$
GeV. More precisely, $|M|$ reaches its maximal allowed value at the point where
the contour for \oh=1 (which has a positive slope with increasing $|\tanb|$)
meets the line \mstau=\mchi\ (which has a negative slope). The location of this
crossing point obviously depends on how large \mstau, $m_P$ and the $P \chi
\chi$ coupling $g_{P \chi \chi}$ are for a given choice of $M$ and \tanb. We
have already mentioned in the previous section that both $|\mu|$ and $m_P$ grow
with increasing $X_2$, eq.(\ref{e26}). Moreover, choices of parameters that
maximize $X_2$ also maximize the reduction of the diagonal elements of the
$\tilde{\tau}$ mass matrix due to the effect of the $\tau$ Yukawa coupling on
the RGE. Finally, increasing $|\mu|$ further reduces \mstau\ by increasing
$\tilde{\tau}_L - \tilde{\tau}_R$ mixing, and reduces $g_{P \chi \chi}$ by
reducing the higgsino component of $\chi$, see eq.(\ref{e6a}). Combinations of
parameters that increase $X_2$ therefore push the $P$ pole to larger values of
$|\tanb|$, and also make it narrower and shallower; at the same time they
strengthen the upper bound on $|\tanb|$ that results from the requirement
$\mstau \geq \mchi$. The combined effect of these changes is to move the
crossing point of the lines \mstau = \mchi\ and \oh = 1 to smaller $|M|$ and
smaller $|\tanb|$.

We already discussed in connection with figs. 5 that in the region of large
$|\tanb|$ one has $A>0$, so that  $X_2$ is smaller for $M<0$ than for $M>0$.
The absolute bound on $|M|$ for given $m$ and $m_t$ is therefore reached if $M$
and \tanb\ are both negative, see fig. 5d. The resulting upper bound on $|M|$
as a function of $m$ is shown in fig. 7 for three different choices of $m_t$.
For $m \leq 125$ GeV the upper bound on $|M|$ only comes from the requirement
$\mstau \geq \mchi$; for these small values of $m$ the constraint (\ref{e34})
is satisfied even for the extreme case $\mchi = \mlr$. Here the maximal allowed
$|M|$ occurs at rather small values of $|\tanb|$, where $\mstau \simeq
m_{\tilde{\tau}_R}$ only depends on $m$ and $M$; for small $m$ the bound on
$|M|$ is therefore almost independent of $m_t$. However, once the condition
$\oh \leq 1$ starts to impose nontrivial constraints on the allowed parameter
space, the bound on $|M|$ comes from the region of large $|\tanb|$ and does
depend quite strongly on $m_t$. For light top quark and thus small $X_2$ (short
dashed curve), the DM constraint reduces $|M|_{\rm max}$ only marginally from
the value of $5.7 m$ that follows from the simple requirement $\mlr > \mchi$.
On the other hand, if the top quark is very heavy (long dashed curve), a
relatively large ratio $|m/M|$ is needed to get sufficiently close to the $P$
pole without reducing \mstau below \mchi. Finally, $|M|_{\rm max}$ also depends
quite sensitively on the chosen combination of signs; for instance, if $M$ and
\tanb\ are both positive one finds $M < $2150 (1210, 620) GeV for $m$ = 500 GeV
and $m_t$ = 110 (140, 170) GeV, respectively.

The rise of the curves in fig. 7 cannot persist indefinetely. Right on the
pole,
for $\mchi = m_P/2$, the annihilation cross section is proportional to
$g^2_{P \chi \chi}/\Gamma_P^2$. The decay width $\Gamma_P$ is proportional to
$m_P$, and thus also to \mchi; moreover, the $P \chi \chi$ coupling decreases
$\propto 1/\mchi$ if $\chi$ is bino--like, see eq.(\ref{e6a}). As a result, the
annihilation cross section for a bino--like LSP on the $P$ pole decreases
$\propto 1/m^4_{\chi}$. We estimate that \oh\ will be larger than 1 even right
on the pole if $\mchi \geq$ 3.5 (1.7) TeV for $m_t$ = 140 (170) GeV. If the
top quark is lighter, one can arrange $\mchi = m_P/2$ even for a mixed LSP,
i.e.
if $|M_1| \simeq |\mu|$. In this case $g_{P \chi \chi}$ is not suppressed by
small mixing angles, which increases the cross section by a factor $\left(
\mchi
/M_Z \right)^2$ compared to the case of a bino--like LSP. The relic density of
a
mixed LSP with $\mchi = m_P/2$ will therefore only exceed 1 if $\mchi > {\cal
O}
(100)$ TeV or so! One can hardly speak of ``weak scale'' supersymmetry if the
{\em lightest} superparticle is thousand times heavier than the weak gauge
bosons; in particular, the model can no longer provide a solution of the
naturalness problem.

It can be argued that finetuning is needed to achieve $|\tanb| \gg 1$, because
this occurs only over a narrow range of values of $A$. Moreover, in minimal
supersymmetric $SU(5)$ the proton decay width increases \cite{54} $\propto
\tan^2 \! \beta$, since in this model proton decay is mostly mediated by the
exchange of superheavy higgsinos, whose Yukawa couplings grow $\propto
|\tanb|$; large values of $|\tanb|$ are then disfavored (although
``accidental'' cancellations might still lead \cite{54} to an acceptable
nucleon lifetime even if $|\tanb| \gg 1$). Fig. 7 therefore also includes
curves (dotted) where we have required $|\tanb| \leq 15$. This rather mild
constraint does not affect the curve for $m_t$ = 110 GeV at all, but for a
heavier top quark it leads to a substantial reduction of $|M|_{\rm max}$. Since
$\mchi \simeq m_P/2$ is no longer possible in this case, the only possibility
to achieve large values of $|M|$ is to choose parameters such that the LSP is
higgsino--like.

We have already seen in fig. 3 that the relic density of such a state is quite
small. By extrapolation of the curves of this figure it is clear that a
higgsino--like LSP is cosmologically safe up to \mchi\ = 2 TeV at least; this
has already been shown in refs. \cite{18} and \cite{19}. The LSP will obviously
only be higgsino--like if $|M_1| \simeq 0.43 |M| \geq |\mu|$, but this
requirement might clash with the constraints imposed by radiative gauge
symmetry breaking, eqs.(\ref{e25}) and (\ref{e26}). Clearly $\mu^2 (Q_0)$ can
be minimized by choosing $A$ such that $X_2$ is minimal, $A \simeq -2.08 M$.
For $\tan^2 \! \beta \gg 1$, eqs.(\ref{e25}) and (\ref{e26}) then give the
following lower bound on $|\mu|$: \beq \label{e37}
\mu^2(Q_0) &\geq m^2 \left[ 0.9 \left( \frac {m_t} {150 \ {\rm GeV}} \right)^2
- 1 \right] \nonumber \\
&+ M^2 \left\{ \left( \frac {m_t} {150 \ {\rm GeV}} \right)^2
\left[ 2.7 -1.04 \left( 1- \left( \frac {m_t} {190 \ {\rm GeV}} \right)^3
\right) \right] - 0.52 \right\}. \eeq
Notice that the coefficient of $M^2$ is positive for $m_t \geq 85$ GeV; it
surpasses $(M_1/M)^2 \simeq 0.18$ for $m_t \geq 95$ GeV, i.e. in almost the
entire experimentally allowed region. The condition for having a higgsino--like
LSP, $|\mu(Q_0)| \leq |M_1|$, is therefore most easily fulfilled if $|M|$
(and thus $|M_1|$) is itself very small. However, the coefficient of $m^2$ in
eq.(\ref{e37}) also turns positive for $m_t > 158$ GeV. For such a heavy top
quark a higgsino--like LSP can therefore {\em not} be realized in minimal
SUGRA. For this reason the lower dotted curve in fig. 7, which is valid for
$m_t = 170$ GeV, is essentially just given by the bound (\ref{e34}) in
the region where DM constraints are relevant and the condition (\ref{e35})
is fulfilled; for $m > 300$ GeV this curve simply follows the $h$ pole,
$\mchi \simeq 0.43 |M| \simeq m_h/2$.

For $m_t$ = 140 GeV (upper dotted curve) a higgsino--like LSP is still
possible provided $|M/m| \leq 0.5$ or so. The actual bound on $|M|$ is
in many cases substantially above this value since for moderately large $m$
the total $\chi \chi$ annihilation cross section is still sufficiently large
even if the higgsino component of $\chi$ is sub--dominant; in particular
the contribution from the \ttbar\ final state plays an important role here.
However, increasing $m$ decreases the sfermion
exchange contribution to the total annihilation cross section; this has
to be compensated by increasing the higgsino content of $\chi$, i.e. by
reducing $|\mu/M|$, which via eq.(\ref{e37}) reduces the bound on $|M/m|$.
As a result the bound on the absolute value of $|M|$ is almost independent
of $m$ for 400 GeV $\leq m \leq$ 1 TeV; for even larger values of $m$ the LSP
can be an almost pure higgsino even for $M >$ 0.5 TeV, and one has
$|M|_{\rm max} \simeq 0.6 m$. Finally, we mention that a further sharpening
of the bound on $|\tanb|$, e.g. requiring $|\tanb| \leq 5$, would suppress
the bound on $|M|$ for $m_t$ = 140 GeV to a value very close to the one for
$m_t$ = 170 GeV. This is because solutions with $\mu^2 \ll m^2$ always
have $\tan^2 \! \beta \gg 1$, unless one has $|A| \gg m$; however,
eqs.(\ref{e25}) and (\ref{e26}) show that large $|A/m|$ require $|\mu|$ to
be larger than $|M_1|$, unless $m_t$ is near its present lower bound \cite{55}
of 91 GeV.

This completes our discussion of possible upper bounds on mass parameters of
the model that can be derived from the requirement $\oh \leq 1$. What about
lower bounds? In principle the bound (\ref{e34}) also implies a lower bound
on the mass of a bino--like LSP if $m$ is fixed within the region allowed
by the limit (\ref{e35}). Recall, however, that these bounds are only valid
if $\chi$ is a nearly pure bino and \mchi\ is not close to a pole. Both these
conditions can be violated quite easily especially if \mchi\ is not large.
Indeed, figs. 5 and 6 show that for most combinations of $m, \ m_t$ and
\tanb\ the lower bound on $|M|$ is determined from laboratory search limits
alone, the exception being the region of small $|\tanb|$ and $m>$ 200 GeV. In
particular, inclusion of the DM constraint does not strengthen the bound
$\mchi > $ 20 GeV that follows \cite{44} from the combination of the gluino,
chargino and neutralino search limits.

On the other hand, we have seen in fig. 6d that the LSP relic density becomes
quite small if $m$ is small, for all experimentally allowed combinations of
the remaining parameters. Imposing an upper bound on $m$ therefore leads to
an upper bound on \oh; conversely, requiring the neutralino relic density to
be larger than some minimal value can give a lower bound on $m$. We have
already
seen that for fixed $m$ the annihilation cross section of a bino--like LSP
becomes small both at very small and at very large \mchi. In the experimentally
allowed \cite{44} region $\mchi \geq 20$ GeV and for $m \leq 140$ GeV,
\oh\ is maximized if $|M|$ is chosen
as large as is allowed by the condition $\mstau \geq \mchi$. Since \mstau\
decreases at large $|\tanb|$, \oh\ will be maximal at small values of
$|\tanb|$ where all three $SU(2)$ singlet sleptons are essentially mass
degenerate. On the other hand, \mstau = \mchi\ allows $M^2 \gg m^2$, which
means
that all other sfermions will be too heavy to contribute significantly to the
annihilation cross section, so that eq.(\ref{en1}) applies. Normalizing the
cross section from numerical results of fig. 1 as before, we find
\be \label{e38}
\oh \leq 0.47 \left( \frac {m} {100 \ {\rm GeV}} \right)^2 + 0.085, \ee
where the constant term comes from the $D-$term contribution to \mlr, which
can be significant for small values of $m$; we have checked numerically that
the true bound deviates from (\ref{e38}) by only 10\% or so. For $m > 140$
GeV the maximum of \oh\ for fixed $m$ is reached if \mchi\ is at its
experimental lower bound; however, in this region eq.(\ref{e38}) already
allows $\oh \geq 1$ anyway.

As discussed above, the bound (\ref{e38}) is saturated if \mchi = \mlr,
which implies \mchi = 2.42 $m$ or $M = 5.7 m$; all sparticle masses would
then be substantially larger than $m$. We have therefore also studied the
question how an upper bound on a physical sparticle mass affects the upper
bound on \oh. We find that fixing the mass of the gluino, of the lighter stop,
or of the lightest neutralino or chargino state does not induce a significant
upper bound on \oh. On the other hand, we have seen repeatedly that $SU(2)$
singlet sleptons have an important effect on the LSP relic density. Since
bino--like LSPs annihilate predominantly via $\tilde{l}_R$ exchange, a light
$\tilde{l}_R$ implies a small neutralino relic density; this has already been
observed by Roszkowski \cite{29}.

If $\chi$ is a pure bino, the annihilation cross section (\ref{en1}) is
minimized and \oh\ is maximized for given \mlr\ if \mchi\ is as small as
experimentally allowed \cite{44}, \mchi\ = 20 GeV. For small \mchi\ and
correspondingly small $|M|$ the mass splitting between sfermions need not
be large; the contributions from all \ffbar\ final states will then have
to be included in eq.(\ref{en1}), properly weighted with the fourth power
of the hypercharge of the exchanged sfermion, and with the sfermion masses
given by eq.(\ref{e20}). The resulting prediction for the maximum of \oh\
as a function of \mlr\ is shown by the dotted curve in fig. 8. The other
curves in this figure are results from numerical scans of the entire allowed
parameter space, using the program MINUIT of the CERN program library.

We see that the extended version of eq.(\ref{en1}) does describe the overall
trend of the numerical results; however, the deviation from the full numerical
bounds can be as large as a factor of 2 here. In particular, for small values
of \mlr\ \oh\ can be substantially {\em larger} than one would expect for a
pure bino. This occurs if $\chi$ is a light, strongly mixed state. Such a light
state exists \cite{39} if $|M|$ and $|\mu|$ are both ${\cal O}(M_Z)$ and
$\pro > 0$. In this case the $SU(2)$ and $U(1)_Y$ components of $\chi$ can have
opposite signs (unlike the familiar case of a photino); moreover, the squared
bino component of this state only amounts to typically 50\%. As a result, its
couplings to charged sleptons are strongly suppressed \cite{39}. On the other
hand, such a state always has sizable higgsino components and thus couples to
gauge and Higgs bosons. For $m_t \geq$ 130 GeV and present experimental bounds
on sparticle masses the suppression of the slepton masses is the more important
effect if $\mlr \leq$ 150 GeV; since the existence of such a light mixed
LSP also implies \cite{39} a rather light chargino the actual numerical value
of the upper bound on \oh\ is in this region largely determined by the LEP
chargino and neutralino search limits (\ref{e27a}) and (\ref{e27c}). For
$\mlr >$ 150 GeV the enhancement of the $Z \chi \chi$ and $h \chi \chi$
couplings over--compensates the suppression of the couplings to sleptons;
in this case choosing $\chi$ to be bino--like does indeed maximize \oh, and
the full numerical result comes out quite close to the simple approximation
based on eq.(\ref{en1}).

The curve for $m_t$ = 110 GeV looks quite different from those for larger
values
of $m_t$. One reason is that for $\mlr \leq$ 75 GeV the Higgs search limits
now force one to choose $|\tanb|$ substantially larger than 1; this increases
the $Z \chi \chi$ coupling, which vanishes for $|\tanb|=1$. In this region
\oh\ is therefore again maximized by choosing $\chi$ to be bino-- or
photino--like. For larger values of \mlr, the constraints imposed by the
Higgs search limits are less severe, since the Higgs boson masses tend to
increase with the overall SUSY breaking scale; for 80 GeV $\leq \mlr \leq$
130 GeV the curve for $m_t$ = 110 GeV is therefore close to those for a heavier
top quark. However, for even larger values of \mlr, \oh\ is maximized for
such small values of $|M/m|$ that radiative gauge symmetry breaking with a
light top quark can only be achieved if $|A/m|$ is quite large, which greatly
reduces the mass of the lighter stop eigenstate; the bound $m_{\tilde {t}_1}
\geq$ 45 GeV then rules out large regions of parameter space, causing the curve
for $m_t$ = 110 GeV to again fall below those for larger values of $m_t$ if
$\mlr >$ 130 GeV.

In all cases we find that the upper bound on \oh\ for light $\tilde{l}_R$
is determined not only by \mlr\ itself, but also by the experimental bounds
on the masses of the other sparticles and Higgs bosons. This dependence is
further illustrated by the long--short dashed curve in fig. 8, which is
valid for $m_t$ = 140 GeV after we impose the (hypothetical) bounds
$m_{\tilde{t}_1, \tilde{\chi}^+} \geq$ 80 GeV and $m_{\tilde g} \geq$ 200 GeV.
This is meant to approximate the bounds that would emerge if LEP200 and the
tevatron fail to find evidence for supersymmetry (other than perhaps light
sleptons). The contraints from searches for Higgs bosons and neutralinos
will also become more severe in the next few years, but the final bounds
depend quite strongly on the energy and luminosity that will be achieved in
the LEP upgrade. The increased bound on the chargino mass excludes a mixed
LSP unless \cite{39} $\mchi \geq$ 40 GeV, which is quite close to the $Z$ pole;
the upper bound on \oh\ is therefore now always saturated if $\chi$ is a
bino-- or photino--like state. The increased lower bound on the gluino mass
implies $\mchi \geq$ 35 GeV for such a state, see eq.(\ref{e21}); this
enhances the annihilation cross section by approximately a factor of 2,
compared to the present bound $\mchi \geq$ 20 GeV. Since $\chi$ is now always
gaugino--like, the bound that can be derived from eq.(\ref{en1}) reproduces the
long--short dashed curve to an accuracy of about 20\%.
\setcounter{footnote}{0}
\section*{5. Summary and Conclusions}
In this paper we have computed the relic density \oh\ of LSPs produced during
the Big Bang, within the framework of minimal $N=1$ Supergravity models with
radiative gauge symmetry breaking. In sec. 2 we briefly described the
calculation of the relic density for a given LSP annihilation cross section.
We then discussed the various contributions to this cross section, including
final states composed of gauge and Higgs bosons that become accessible for a
``heavy'' LSP, $\mchi > M_W$. Expressions for all these cross sections are
given in the Appendix, for a general neutralino eigenstate; to our knowledge
such a complete list does not exist in the literature. In sec. 2b we discussed
the qualitative features of the most important contributions to the
annihilation
cross section for the important special cases that the LSP is an almost pure
higgsino or bino. In particular, we pointed out that, even though the higgsino
component of a bino--like LSP vanishes as $\mchi \to \infty$, the matrix
elements for $\chi \chi \to VV$ and $\chi \chi \to V H$ remain {\em finite},
where $V = W,Z$ and $H$ is a Higgs boson. This is because of the enhanced
production of longitudinal gauge bosons; the same result can also be derived
from the equivalence theorem.

In sec. 2c we briefly summarized the constraints imposed on the particle
spectrum by the assumption of minimal supergravity (SUGRA). The relations
between the masses of the gauginos and those of the superpartners of the light
quarks and leptons are by now well known, and have been included in several
previous analyses \cite{20,21,22,23}. However, SUGRA also implies relations
between the mass of the top quark and the soft breaking parameters on the one
hand, and the masses of the Higgs bosons as well as the supersymmetric
Higgs(ino) mass parameter $\mu$ on the other; these relations had previously
only been included \cite{25,24} for the case that the SUSY breaking gaugino
mass
$M$  is much larger than the scalar mass $m$ at the GUT scale. We saw in our
numerical examples of sec. 3 that these relations can have large effects on the
relic density. In particular, we found that the coupling of a bino--like LSP to
gauge and Higgs bosons is usually suppressed for a heavy top quark and/or large
$m$, since increasing $m_t$ or $m$ tends to increase $|\mu|$, which in turn
reduces the higgsino component of $\chi$. This strongly affects \oh\ both in
the pole region ($\mchi \simeq M_Z/2$) and in the threshold region ($\mchi
\simeq M_Z$). Moreover, as already pointed out in ref.\cite{DN1}, \oh\ is
greatly reduced if the ratio \tanb\ of the vevs of the two Higgs bosons is
large, since this implies large $b$ and $\tau$ Yukawa couplings, and thus a
light pseudoscalar Higgs boson $P$ and reduced masses for the light $\tilde{b}$
and $\tilde{\tau}$ eigenstates.

The relations between particle masses and neutralino couplings imposed by SUGRA
also imply that the region of parameter space where the LSP density can lead to
a flat universe (0.25 $\leq \oh \leq$ 1), as predicted by inflationary models,
has a very complicated shape, since the $\chi \chi$ annihilation cross section
depends on {\em all} input parameters; the same is true for the region where
LSPs can at least provide the DM halo of galaxies ($\oh \geq 0.025$).
Nevertheless it is clear from the results of sec. 3 that a cosmologically
interesting relic density is obtained quite naturally, provided that $\chi$ is
gaugino--like and the SUSY breaking parameter $m$ is not too small. If $\chi$
is higgsino--like and $\mchi > M_W$, $\chi \chi$ annihilation into pairs of
gauge bosons is so strong that cosmologically interesting values of \oh\ only
occur \cite{18,19} for $\mchi >$ 0.5 TeV. Since in minimal SUGRA the gluino as
well as most squarks are at least 6 times heavier than the LSP, requiring
$\mchi >$ 0.5 TeV leads to quite severe fine tuning \cite{31,16} and is thus
unattractive. The $\chi \chi$ annihilation cross section for higgsino--like
$\chi$ can be quite small below the $WW$ threshold if the gaugino mass $|M|$ is
large; however, using results of ref.\cite{33} we estimated that in this case
co--annihilation of $\chi$ with the next--to--lightest neutralino $\chi'$ will
suppress \oh\ to a value well below 0.25, since the off--diagonal $Z \chi
\chi'$ coupling is large, and since co--annihilation can proceed from an
$s-$wave initial state. We therefore conclude that the higgsino does not make a
very attractive DM candidate in minimal SUGRA models.

Unfortunately we saw in sec. 4 that the complicated shape of the region of
parameter space that yields $\oh \leq$ 1 makes it difficult to derive
interesting upper bounds on mass parameters from the requirement that relic
neutralinos do not overclose the universe. In particular, $m$ can be almost
arbitrarily large, and thus all sfermions and most Higgs bosons can be very
heavy, if $\mchi \simeq M_Z/2$ or $\mchi \simeq m_h/2$, where $h$ denotes the
light scalar Higgs boson. In this case the gluino, the lighter chargino and the
next--to--lightest neutralino would still lie in the region accessible to the
next generation of accelerators ($m_{\tilde g} \leq 500$ GeV, $m_{\tilde
{\chi}^+, \chi'} \leq$ 150 GeV). However, if $|\tanb|$ is large, $m_P$ can be
greatly reduced so that \mchi\ can be close to $m_P/2$. Since $P$ exchange
mediates $\chi \chi$ annihilation from an $s-$wave state, \oh\ is greatly
suppressed in this case, allowing both $m$ and $|M|$ to be well beyond 1 TeV
without getting in conflict with cosmology. If the top quark is not very heavy,
$m_t \leq 155$ GeV, cosmologically safe solutions with very large $m$ and $|M|$
also exist for small $|\mu|$, so that the LSP is higgsino--like. In this case
one would still have a ``light'', almost degenerate $SU(2)$ doublet of
higgsinos, but all other sparticles would be very heavy. Since the mass
splitting could be as small as ${\cal O}(1$ GeV), such light higgsinos would
probably be very difficult to observe even at $e^+ e^-$ colliders \cite{56}.
This scenario is therefore very similar to the case of a very large \mchi\ as
far as collider experiments are concerned. We have to conclude that the
requirement $\oh \leq$ 1 does not strictly exclude the possibility that
sparticle masses are well beyond the reach even of the planned supercolliders.

On the other hand, interesting upper bounds can be obtained for the ``generic''
case of a bino--like LSP away from poles. We found that in this case the total
annihilation cross section is dominated by the $l^+ l^-$ final state produced
via $\tilde{l}_R$ exchange ($l = e,\mu,\tau$). Contributions from $\tilde{l}_L$
exchange and $\tilde{q}$ exchange are suppressed by the larger masses and
smaller hypercharges of these sfermions, while the contribution from $Z$
exchange is suppressed by the small $Z \chi \chi$ coupling. We mentioned
earlier that the $VV$ and $Zh$ final states also contribute with full gauge
strength if the LSP is sufficiently heavy, but the relevant hypercharge here is
the one of the Higgs bosons, leading to a suppression factor of 1/16 compared
to the $\tilde{l}_R$ exchange contribution. This allowed us to derive the very
simple and yet quite accurate expression (\ref{en1}) for the total annihilation
cross section, leading to the analytically derived bounds (\ref{e35}) and
(\ref{e36}). In particular, we find for this case an upper bound $m_{\tilde g}
\leq $ 2 TeV, which is only slightly less stringent than the bound derived
\cite{31,16} from the requirement that fine tuning should occur at most at the
10\% level; the corresponding bound on \mchi\ is considerable more stringent
than the one that follows under similar assumptions \cite{18} in a more general
SUSY model. Our bound $m \leq$ 300 GeV is potentially even more interesting,
since this value is {\em below} the one derived \cite{31,16} from fine tuning
arguments; together with the bound on $m_{\tilde g}$ it would virtually
guarantee that at least the sleptons, the light chargino and the
next--to--lightest neutralino should be observed at a TeV $e^+ e^-$
supercollider. However, we have to remind the reader that there are several
ways to evade these bounds.

As first pointed out in ref.\cite{21} and also emphasized in refs.\cite{22,23},
the LSP relic density can only lead to a flat universe with $0.5 \leq h \leq 1$
if $m$ is not too small. We quantified this in the simple relation (\ref{e38}),
which shows that $\oh \geq 0.25$ is only possible for $m \geq 40$ GeV. (In
ref.\cite{23} the more stringent bound $m \geq 100$ GeV has been found, but
this is only valid for $\mchi \leq M_W$.) Furthermore, requiring $\oh \geq
0.25$ also implies a lower bound on the mass of $SU(2)$ singlet sleptons. The
exact value of this bound depends on the top mass, as well as on the bounds on
the masses of other sparticles and Higgs bosons, but it is always close to 100
GeV, see fig. 8. We thus see that requiring 0.25 $\leq \oh \leq$ 1 determines
\mlr\ to be within a factor of 2 of 200 GeV, {\em if} the LSP is gaugino--like
and not near a pole. Unfortunately, in this case LEP200 will fail to discover a
slepton. This has already been pointed out in ref.\cite{29} for a light LSP,
but at least within the framework of SUGRA models this conclusion also holds
$\mchi \geq M_W$. (Of course, we need $\mchi \leq \mlr$ always.) On the other
hand, the gaugino mass parameter $M$ is only poorly determined by the
requirement that \oh\ lies in the cosmologically interesting range, since the
annihilation cross section is almost independent of $M$ over a wide range, see
figs. 1. In particular, the gluino, the squarks, the lighter chargino and at
least one neutralino could all lurk ``just around the corner'', but their
masses could also lie in the range that can only be covered by supercolliders.

Our overall conclusion is that, while limits on the relic neutralino density
allow to rule out large regions of parameter space, they do not allow to derive
upper bounds on sparticle masses which are both interesting for experiments at
existing or planned colliders and valid for all combinations of the other
parameters. On the other hand, results for the perhaps most natural case of a
gaugino--like LSP do indicate that sparticle masses should lie in the range to
be covered by planned $e^+ e^-$ and $pp$ supercolliders. Moreover, if the dark
matter in our galaxy does indeed consist of neutralinos, one would expect their
mass to lie within the range of sensitivity \cite{57} of next--generation
direct search experiments and \cite{58} of experiments looking for neutralino
annihilation in the sun, although a strict lower bound on the expected signal
size is again difficult to derive. Cosmological arguments can therefore
supplement and lend support to direct SUSY searches at collider experiments.

\subsection*{Acknowledgements}
We thank K. Hikasa for sharing his expertise in the calculation of helicity
amplitudes with us, as well as for many useful discussions. M.M.N. thanks the
theory group at KEK, and M.D. the theory group at SLAC and the Institute of
Particle Physics Phenomenology in Madison, Wisconsin, for their kind
hospitality.
\renewcommand{\theequation}{A.\arabic{equation}}
\setcounter{equation}{0}
\section*{Appendix A: Matrix elements}
\newcommand{\iso}{\mbox{$^1S_0$}}
\newcommand{\tpo}{\mbox{$^3P_0$}}
\newcommand{\tpi}{\mbox{$^3P_1$}}
\newcommand{\tpii}{\mbox{$^3P_2$}}
\newcommand{\li}{\mbox{$\lambda_i$}}
\newcommand{\lf}{\mbox{$\lambda_f$}}
\newcommand{\barh}{\mbox{$\bar h$}}
\newcommand{\dli}{\mbox{$\delta_{\lambda_i,0}$}}
\newcommand{\mbarh}{\mbox{$(-1)^{{\bar h}-\frac{1}{2}} $}}
\newcommand{\mbrh}{\mbox{$(-1)^{{\bar h}+\frac{1}{2}} $}}
\newcommand{\dif}{\mbox{$d^J_{\lambda_i,\lambda_f}$}}
\newcommand{\ols}{\mbox{${O_{0j}^ L}^2$}}
\newcommand{\ors}{\mbox{${O_{0j}^R}^2 $}}
\newcommand{\olr}{\mbox{$O_{0j}^{L}O_{0j}^R$}}
\newcommand{\pj}{\mbox{$1+{R_j^+}^2-R_W^2$}}
\newcommand{\olsp}{\mbox{${O_{0j}^{''L}}^2$}}
\newcommand{\betb}{\mbox{$\bar\beta_f$}}
\newcommand{\jp}{\mbox{$J_j^{'}+J_j^{''}$}}
\newcommand{\jm}{\mbox{$J_j^{'}-J_j^{''}$}}
\newcommand{\kp}{\mbox{$K_j^{'}+K_j^{''}$}}
\newcommand{\km}{\mbox{$K_j^{'}-K_j^{''}$}}
\newcommand{\op}{\mbox{$O_{0j}^R+O_{0j}^L$}}
\newcommand{\om}{\mbox{$O_{0j}^R-O_{0j}^L$}}
\newcommand{\qp}{\mbox{$Q_{0j}^{'R}\sin\!\beta+Q_{0j}^{'L}\cos\!\beta$}}
\newcommand{\qm}{\mbox{$Q_{0j}^{'R}\sin\!\beta-Q_{0j}^{'L}\cos\!\beta$}}
\newcommand{\qlr}{\mbox{$Q_{0j}^{'L}Q_{0j}^{'R}$}}
\newcommand{\qlpr}{\mbox{${Q_{0j}^{'L}}^2+{Q_{0j}^{'R}}^2$}}
\newcommand{\qlmr}{\mbox{${Q_{0j}^{'L}}^2-{Q_{0j}^{'R}}^2$}}
\newcommand{\xpw}{\mbox{${X_{a0}^{'}}^2+{W_{a0}^{'}}^2$}}
\newcommand{\xmw}{\mbox{${X_{a0}^{'}}^2-{W_{a0}^{'}}^2$}}
\newcommand{\xw}{\mbox{$X_{a0}^{'}W_{a0}^{'}$}}
\newcommand{\pjp}{\mbox{$1+{R_j^{+}}^2$}}
\newcommand{\xlrz}{\mbox{$X_{a0}^{'}\leftrightarrow Z_{a0}^{'}$}}
\newcommand{\wlry}{\mbox{$W_{a0}^{'}\leftrightarrow Y_{a0}^{'}$}}
\newcommand{\hlrl}{\mbox{$P_{1}\leftrightarrow P_{2}$}}
Since we are interested in the nonrelativistic limit of the $\chi \chi$
annihilation cross section, we employ the partial wave formalism. In this
formalism the helicity amplitude for the process
 \be\label{a1}
 \chi (h)+\chi ( \barh ) \rightarrow X(\lambda_X ) +Y( \lambda_Y )
\ee
 ($h,\barh,\lambda_X$ and
  $\lambda_Y$ are the helicities of the corresponding particles)
 is expanded as follows:

 \be \label{a2}
  T=\sum^{\infty}_{L=0} \sum^{1}_{S=0} \sum^{L+S}_{J=|L-S|}
 A(^{2S+1}L_J) {\cal P}(^{2S+1}L_J) \dif.
\ee

Here the reduced partial wave amplitude $A$ describes annihilation from an
initial state with definite spin $S$ and orbital angular momentum $L$, and thus
also with definite $C$ and $P$ quantum numbers. The spin projectors ${\cal P}$
depend only on $h$ and \barh, and the angular dependence is contained in the
$d$ functions \dif. $\li=h-\barh$ and $\lf=\lambda_X-\lambda_Y$ are the
differences of the helicities of the initial and final particles, respectively.
Because our initial state consists of two identical Majorana fermions, we only
need to consider initial states with $C=1$. Furthermore, since we want to
expand the total annihilation cross section only up to ${\cal O}(v^2)$, only
annihilation from $s-$ and $p-$wave initial states has to be included.
Altogether we thus find that we need to include only the contributions from
the \iso, \tpo, \tpi\ and \tpii\ initial states; explicit expressions
for the relevant ${\cal P}$ can be found in ref.\cite{samurai}.
The annihilation cross section is then given by
\be\label{a5}
\sigma(\chi \chi \rightarrow XY)v=\frac{1}{4}
\frac{ \bar\beta_f}{8\pi s {\rm S}}
\left\{ |A(\iso )|^2+\frac{1}{3}
\left[|A(\tpo)|^2+|A(\tpi )|^2+|A(\tpii )|^2 \right] \right\}.\ee
Here $v$ is the the relative velocity of initial neutralinos and $s$ is the
total energy.  S is a symmetry factor which is 2 when $X=Y$. The summation over
the final helicities is implicit in  this equaiton. Finally, the kinematical
factor \betb\ is given by \be \label{beta}
\bar\beta_f=\sqrt{1-2\frac{m_X^2 +m_Y^2}{s}+\frac{(m_X^2-m_Y^2)^2}{s^2}}.
\ee

In this Appendix we list the helicity amplitudes $A(^{2S+1}L_J)$ for all
two--body final states accessible to $\chi \chi$ annihilation in leading order
in perturbation theory. We first list some couplings which appear in many
matrix elements:

\ben\label{a50}\beq
O_{0j}^{L}&=-\frac{1}{\sqrt{2}}N_{04}V_{j2}+N_{02}V_{j1}\\
O_{0j}^R&=\frac{1}{\sqrt{2}}N_{03}U_{j2} +N_{02}U_{j1}\\
Q_{0j}^{'L}&=N_{04}V_{j1}+\frac{1}{\sqrt{2}}
(N_{0j}+N_{01}\tan\theta_W)V_{j2}\\
Q_{0j}^{'R}&=N_{03}U_{j1}-\frac{1}{\sqrt{2}}
(N_{02}+N_{01}\tan\theta_W)U_{j2}\\
O_{0j}^{''L}&=-\frac{1}{2} N_{03}N_{j3}+\frac{1}{2} N_{04}N_{j4}\\
Q_{0j}^{''}&=\frac{1}{2}\left[N_{03}(N_{j2}-\tan\!\theta_W N_{j1})
+( 0\leftrightarrow j)\right] \\
S_{0j}^{''}&=\frac{1}{2}\left[N_{04}(N_{j2}-\tan\!\theta_W N_{j1})
+( 0\leftrightarrow j)\right]
\eeq\een
The expressions for $O^L, \ O^R$ and $O''^L$ are the same as in ref.\cite{6};
our definitions for $Q'^L, \ Q'^R, \ Q''$ and $S''$ differ slightly from
those of ref.\cite{38}. $U$ and $V$ are the matrices that diagonalize the
chargino mass matrix ${\cal M}^{\pm}$, which is given by \cite{6}:
\be\label{a51}
{\cal M}^{\pm} = \left(\begin{array}{cc}
M_2 & M_W\sqrt{2} \sin\beta\\
M_W\sqrt{2}\cos\beta & \mu
\end{array}\right).
\ee
The matrices $U$ and $V$ can be chosen to be real:
\be\label{a52}
U=\left(\begin{array}{cc}
\cos\phi_- &\sin\phi_-  \\
-\sin\phi_- &\cos\phi_-
\end{array}\right),\ \ \
V=\left( \begin{array}{cc}
\cos\phi_+ &\sin\phi_+    \\
-\sin\phi_+ &\sin\phi_+
\end{array}
\right)
\ee
In the limit where either $|M_2|$ or $|\mu|$ is much larger than $M^2_W$, the
chargino mass eigenstates are almost identical to the current states.
The (small) mixing angles then become:
\ben\label{a53}\beq
\phi_-&=\frac{\sqrt{2}M_W}{M_2^2-\mu^2}(M_2\cos\beta+\mu\sin\beta);\\
\phi_+&=\frac{\sqrt{2}M_W}{M_2^2-\mu^2}(M_2\sin\beta+\mu\cos\beta).
\eeq\een
The $N_{ij}$ in eqs.(\ref{a50}) are elements of the matrix that diagonalize the
neutralino mass matrix, eq.(\ref{e4}); as already discussed in sec. 2b, it also
becomes almost diagonal if $|M_2|$ or $|\mu|$ is very large, see
eqs.(\ref{e6}). Unlike refs.\cite{6} and \cite{38}, we do not require the
chargino and neutralino masses to be positive; $N_{ij}$ can therefore also
chosen to be real. We have checked by explicit calculation that this still
preserves all relative signs in our amplitudes, provided we keep the sign of
the fermion masses everywhere. Finally, the suffix zero in eqs.(\ref{a50})
refers to the lightest neutralino, $\chi$.

The coupling of the neutral Higgs bosons to neutralinos can be expressed in
terms of $Q''$ and $S''$ \cite{38}:
\ben\label{a5a}\beq
T_{P0j}=& -\sib Q_{0j}^{''} + \cb S_{0j}^{''}; \label{a5a1}\\
T_{10j}=& -\cos\!\alpha Q_{0j}^{''} + \sin\!\alpha S_{0i}^{''};\\
T_{20j}=&  \sin\!\alpha Q_{0j}^{''} + \cos\!\alpha S_{0i}^{''}; \\
\eeq\een
Here the labels $P$, 1, and 2 refer to the pseudoscalar, and the heavier and
lighter neutral scalar; $\alpha$ in eqs.(\ref{a5a}) is the mixing angle of the
neutral scalar Higgs bosons as defined in ref.\cite{38}, including
leading one--loop radiative corrections.

Finally, we define some kinematical quantities and propagators:
\ben\label{a5b}\beq
\Delta^2&=\frac{m_X^2+m_Y^2}{2 m^2_{\chi}} \label{a5b2};
\\
P_I&=1+R_I^2 -\frac{R_X^2 +R_Y^2}{2}; \label{a5b1} \\
R_X&=m_X/m_{\chi}, \ \ R_Y=m_Y/m_{\chi}, \ \  R_I =m_I/m_{\chi}. \label{a5b3}
\eeq\een
Here, $m_I$ is the mass of an exchanged particle, and $P_I$ the $v \to 0$
limit of the inverse of the corresponding $t$ or $u$ channel propagator.

We are now ready to list the partial wave amplitudes of the contributing
processes. As mentioned above, we only need the nonrelativistic limit of
the annihilation cross section, i.e. its expansion in powers of $v$ up to
and including terms of ${\cal O}(v^2)$. However, this expansion breaks down
\cite{33} in the vicinity of poles and thresholds, since there the ``higher
order'' terms can actually diverge. In our expressions below we therefore
only include those terms of order $v^2$ that result from the expansion
of {\it well--behaved} functions of $v$. Specifically, we include terms
that result from the expansion of $t-$ and $u-$channel propagators, as well
as terms that result from the calculation of spinors or
bosonic wave functions. On the other hand, we do {\em not} expand the
kinematical function \betb, nor $s-$channel propagators. Of course,
far away from the threshold or pole, these terms can be expanded
safely. This can be incorporated into our matrix elements by the
following substitutions:\footnote{Far away from the pole, the propagator
can be taken to be real.} \ben \label{subs} \beq
\betb & \to \betb(v=0) + \frac {v^2} {8 \betb(v=0)} \left[ \Delta^2 -
\frac {\left( m_X^2 - m^2_Y \right)^2} {8 m^4_{\chi}} \right]; \label{subs1} \\
\frac{1}{4-R_I^2} & \to \frac{1}{4-R_I^2} \left( 1 - \frac {v^2} {4-R_I^2}
\right). \label{subs2} \eeq \een
Clearly, these substitutions should only be used in the ${\cal O}(v^0)$
terms of the \iso\ amplitudes. Because of the suppression factor $3/x_F$ in
the expression (\ref{e3}) for \oh, as well as the smallness of the
the numerical factors in eq.(\ref{subs1}), the numerical effect of these
substitutions is only sizable in cases where the expansion itself can no
longer be trusted; in this case only the much more complicated methods
described in ref.\cite{33} give reliable results. Since a more careful
treatment
of poles and thresholds will not change our conclusions, we do not pursue
this avenue here. It should be noted that the ${\cal O}(v)$ terms that
result from the expansion of $t-$ and $u-$channel propagators can, e.g.,
change the annihilation cross section into \ffbar\ final states by a
factor of two; fortunately the ${\cal O}(v)$ terms are always regular.

We list contributions with different final state helicities
separately; all these contributions have to be added incoherently, as
shown in eq.(\ref{a5}).

\subsubsection*{1) $\chi \chi \rightarrow  W^-(\lambda)W^+(\bar\lambda) $}
This final state receives contributions from the exchange of the two chargino
eigenstates (labelled by $j$) in the $t-$ or $u-$ channel, as well as from the
exchange of the two neutral scalar Higgs bosons (labelled by $i$) as well
as the $Z$ boson in the $s-$channel. In the following expressions, summation
over subscripts ($i, \ j, \dots$) that appear more than once is always
understood.

\ben\label{a6}\beq
 &A \left(\iso\right): \  \lf=0,\  \lambda=\pm 1  \nonumber\\
 &\ \ \ \ 2\sqrt{2} \betb g^2\frac{\ols+\ors}{P_j}
  +\sqrt{2} v^2 \betb g^2 \left\{
 \frac{2}{3}\frac{R_j^+}{ P_j^2}\olr\right.
 \nonumber\\
 &\ \ \ \ \ \ \ \ \ \ \ \ \left.  +\frac{\ols+\ors}{P_j}
 \left[ \frac{1}{4}-\frac{4}{3P_j}
 +\frac{2\bar{\beta}_f^2}{3 P_j^2 } \right]
 \right\} \\
&A \left( \tpo \right) : \  \lf=0, \lambda=0
\nonumber\\
&\ \ \ \
 \frac{\sqrt{6}v g^2}{R_W^2}  \left\{ -\frac{4}{3}\frac{\ols+\ors}{P_j}
 +\frac{4\olr R_j^+}{P_j}\left[1-\frac{2}{3P_j}\right]\right\}
\nonumber\\ &\ \ \ \
 +\sqrt{6} v g^2\left\{
 \frac{\ols+\ors}{P_j} \left[ 1-\frac{2\bar{\beta}_f^2}{3P_j }\right]
   -\frac{2\olr R_j^+}{P_j}
   \left[ 1-\frac{4}{3P_j} \right] \right\}
\nonumber\\ &\ \ \
\ \ \ \ \ \ -\frac{\sqrt{6} v  (1+\bar{\beta}_f^2) g^2 F_i}
{(4-R^2_{H_i} + i G_{H_i} ) R_W}
\\
      &\  \lf=0, \lambda=\pm 1 \nonumber\\ &\ \ \ \
  \sqrt{6} v g^2 \left\{
  \frac{\ols+\ors}{P_j}
 \left[\frac{1}{3}-\frac{2\bar{\beta}_f^2}{3P_j}\right]
 -\frac{2\olr R_j^+}{P_j}\right\}
 +\frac{\sqrt{6} v g^2 R_W F_i}{4-R_{H_i}^2 + i G_{H_i}} \\
&A \left( \tpi \right) :\  |\lf|=1,\nonumber\\ &\ \ \ \
  \frac{2 v  \bar{\beta}_f^2\lf g^2} {R_W} \left\{
 \frac{\ols+\ors}{P_j}\left[ 1-\frac{1}{P_j}\right]
 -\frac{2\olr R_j^+}{P^2_j}\right\}\nonumber\\
 &\ \ \ \ \ \ \ \ \ +2v \betb g^2\frac{\ols-\ors}{R_W P_j}
 \left[2-\frac{\bar{\beta}_f^2}{P_j}\right]
 -\frac{8v\betb g^2 O_{00}^{''L}}
{R_W (4-R_Z^2)}
 \\
      &\ \lf=0, \lambda= 0 \nonumber\\&\ \ \ \
\ \ \  \frac{2 v \betb}{R_W^2 P_j} (3-\bar{\beta}_f^2)g^2(\ols-\ors)
-\frac{4 v \betb g^2O_{00}^{''L}}{(4-R_Z^2)R_W^2} (3-\bar{\beta}_f^2)
 \\
      &\ \lf=0, \lambda= \pm 1 \ \ \
 \frac{2v \betb }{P_j}g^2( \ols-\ors)
 -\frac{4v \betb g^2 O_{00}^{''L}}{4-R_Z^2}\\
&A\left( \tpii \right) : \ |\lf|=2 \ \ \ \
 -\frac{2\sqrt{2} v}{P_j} g^2(\ols+\ors)
\\
       & |\lf|=1 \nonumber\\ &\ \ \ \
\frac{2 vg^2}{R_W}\left\{ \frac{-{R_j^+}^2}{P_j^2}(\ols+\ors)
 +\frac{2\olr R_j^+}{P^2_j}{\betb}^2 +\lf{\betb}^3\frac{\ols-\ors}{P^2_j}
\right\}
\\
       &\lf=0,\lambda=\pm 1 \nonumber\\ &\ \ \ \
\frac{2 v g^2}{\sqrt{3}} (\ols+\ors) \frac{1-{R_j^+}^2-R_W^2}{P_j^2}\\
       &|\lf|=0,\lambda=0\nonumber\\ &\ \ \ \
\frac{4v g^2}{\sqrt{3} R_W^2}
\left\{-\frac{\ols+\ors}{P_j}\left[1-\bar{\beta}_f^2\frac{R_W^2}{P_j}\right]
 +4\bar{\beta}_f^2\frac{\olr R_j^+}{P_j^2} \right\}
\eeq\een

Here, the $R$ are the rescaled masses of exchanged or final state particles,
see eq.{\ref{a5b3}), and we have introduced the rescaled widths \be
G_{H_i} = \frac {\Gamma_{H_i} m_{H_i} } {m^2_{\chi}}. \ee
The definition of the $P_j$ is as in eq.(\ref{a5b1}),
and the $F_i$ are given by \cite{38}:
 \be\label{a7}
F_1= \cos(\beta-\alpha)T_{100},\ \ \
 F_2=  \sin(\beta-\alpha) T_{200}
\ee

\subsubsection*{ 2) $\chi \chi \rightarrow  Z(\lambda)Z(\bar{\lambda}) $}
Here the contributing diagrams are very similar to those of the $W^+W^-$
final state, except that the $Z$ exchange contribution is absent. Moreover,
the subscript $j$ now labels the 4 neutralino eigenstates. The nonvanishing
partial wave amplitudes can be written as:
\ben\label{a9}\beq
 & A\left(\iso\right):\  \lf=0\  \lambda=\pm 1\nonumber\\ & \ \ \
  \frac{4\sqrt{2}\betb g_Z^2\olsp}{P_j}
  +2\sqrt{2}  v^2 \betb g_Z^2\olsp\left\{
 -\frac{R_j}{3 P_j^2}
 \right.
\nonumber\\ &\left.\ \ \ \ \ \ \ \ \ \
  +\frac{1}{P_j}\left[ \frac{1}{4}-\frac{4}{3P_j}
 +\frac{2\bar{\beta}_f^2}{3 P_j^2 } \right]
 \right\} \\
 &A\left(\tpo\right) :\  \lf=0, \lambda=0 \nonumber\\ &\ \ \ \
\frac{4\sqrt{6}v g_Z^2}{R_Z^2} \olsp
\left\{ -\frac{2}{3P_j}
-\frac{R_j}{P_j}\left[1-\frac{2}{3P_j}\right]\right\}
\nonumber\\ &\ \ \ \ \ \ \
+2\sqrt{6} v g_Z^2\olsp\left\{
\frac{1}{P_j} \left[ 1-\frac{2\bar{\beta}_f^2}{3P_j }\right]
+\frac{R_j}{P_j}
\left[ 1-\frac{4}{3P_j} \right] \right\}
\nonumber \\
&\ \ \ \ \ \ \ \ \ \ \ \ \ \ \ \ \ \
-\frac{\sqrt{6} v g^2 (1+\bar{\beta}_f^2) F_i}{(4-R_{H_i}^2 + i G_{H_i}) R_W}
\\
      &\  \lf=0, \lambda=\pm 1
\nonumber\\ &\ \ \ \
 2 \sqrt{6} v g_Z^2 \olsp\left\{
  \frac{1}{P_j}
 \left[\frac{1}{3}-\frac{2\bar{\beta}_f^2}{3P_j}\right]
 +\frac{R_j}{P_j}
 \right\}
 +\frac{\sqrt{6} v g_Z^2 R_W F_i}{4-R_{H_i}^2 + i G_{H_i} }\\
&A\left(\tpi\right) :\  |\lf|=1,\ \ \ \
 \frac{ 4v\bar{\beta}_f^2\lf g_Z^2\olsp} {R_Z}\left\{
 \frac{1}{P_j}\left[ 1-\frac{1}{P_j}\right]
 +\frac{R_j}{P_j}\right\}
 \\
&A\left(\tpii\right) :\ |\lf|=2 \ \ \
 -\frac{4\sqrt{2} v g_Z^2}{P_j}  \olsp
\\
       & |\lf|=1 \ \ \ \ \ \ \ \ \
-\frac{4vg_Z^2 }{R_Z} \olsp\left\{ \frac{R_j^{2}}{P_j^2}
 +\frac{R_j}{P_j}\bar{\beta}_f^2
\right\}
\\
       &|\lf|=0,\lambda=\pm 1 \ \ \
\frac{4v g_Z^2}{\sqrt{3}} \olsp \frac{1-R_j^2-R_Z^2}{P_j^2}\\
       &|\lf|=0,\lambda=0\ \ \
-\frac{8 v g_Z^2 }{\sqrt{3} R_Z^2}\olsp
\left\{\frac{1}{P_j}\left[1-\frac{R_Z^2\bar{\beta}_f^2}{P_j}\right]
 +\frac{ 2R_j\bar{\beta}_f^2}{P_j^2} \right\}
\eeq\een
In eqs.(\ref{a9}) we have used the usual notation \be
g_Z = \frac {g}{\cos \! \theta_W}. \ee
\subsubsection*{3) $\chi \chi \rightarrow  Z(\lambda) H_a $}
This final state receives contributions from the exchange of the four
neutralinos (labelled by $j$) in the $t-$ or $u-$channel, as well as from
the $s-$channel exchange of the $Z$ boson as well as of the pseudoscalar
Higgs bosons $P$. The result is:
\ben\label{a11}\beq
A \left(\iso \right) : \ &\lambda = 0 \nonumber\\
&-\frac{2\sqrt{2}\betb}{R_Z}
g g_Z \left\{ \frac{2J_j(R_j-1)}{P_j} +\frac{O}{R_Z\cw}
 - \frac{4L}{4-R_P^2+iG_P}\right\}
\nonumber\\
&- v^2 \frac{\sqrt{2}\betb}{R_Z}g g_Z
\frac{J_j}{P_j}\left\{\frac{1}{2}(R_j-5) - \frac{2(R_j-1)}{P_j}
+\frac{4(R_j-1)}{3 P_j^2} \betb^2\right.
\nonumber\\
&\left. \ \ \ \ \ \ \ \ \ \ \ \ \ +( 2-\Delta^2 )\frac{2}{3P_j}
\right\}
\nonumber\\
&+v^2 \frac{\sqrt{2}\betb}{R_Z}g g_Z
\left\{ \frac{3 L}{4-R_{P}^2+i G_P}
 -\frac{O}{4R_Z\cw}\right\}
\\
A \left(\tpi \right) :& \lambda=\pm 1\nonumber\\
&   -4v g g_Z \frac{J_j }{P_j^2} \left[R_j^2-\frac{\delta^4}{4} \right]
 -4vg g_Z \frac{J_j R_j}{P_j}+2v g_Z^2  O \frac{R_Z}{4-R_Z^2}
\\
&\lambda=0
\nonumber\\
&-2\frac{v}{R_Z}(1+\frac{\delta^2}{2} ) g g_Z
\left\{ 2(1+R_j)\frac{J_j}{P_j} - \frac{R_Z}{\cw (4-R_Z^2)}O \right\}
\\
A \left( \tpii\right) : &\lambda=\pm 1
\ \ \ \ \ -4\lambda v \betb^2 g g_Z\frac{J_j}{P_j^2}
 \eeq\een
In addition to $\Delta$, which has already been defined in
eq.(\ref{a5b2}), we have introduced the following quantities:
\ben\label{a13}\beq
G_P = &\frac{\Gamma_P m_P}{m^2_{\chi}}; \\
\delta^2=&\frac{R_Z^2-R_{H_a}^2}{2}; \\
J_j=&-O_{0j}^{''L} T_{a0j}\ \ \ \ \ {\rm for} \ H_a;   \\
L=
& \frac{1}{2}\sin (\alpha-\beta)T_{P0j} \ \ \ \ \  {\rm for}\  H_1;
\\
& \frac{1}{2}\cos(\alpha-\beta)T_{P0j} \ \ \ \ \ {\rm for}\  H_2; \\
O= & O_{0j}^{''L} \sin (\alpha - \beta) \ \ \ \ \ {\rm for} \ H_1; \\
   & O_{0j}^{''L} \cos (\alpha - \beta) \ \ \ \ \ {\rm for} \ H_2.
\eeq\een

\subsubsection*{4) $\chi \chi \rightarrow  Z(\lambda) P$}
The contributing diagrams are very similar to those of the $ZH$ final state,
except that the single $P$ exchange diagram has to be replaced by
$H_1, \ H_2$ exchange diagrams; notice that these scalar Higgs bosons cannot
be on--shell, so that we need not introduce complex propagators here.
\ben\label{a14}\beq
A \left(\tpo\right) :& \lambda=0\nonumber\\
 &\ \  4\sqrt{6}\frac{v \betb }{R_Z}g g_Z
 \left\{
 \left[1+\frac{\delta^2}{2}\right] \frac{K_jR_j}{3P_j^2}
-\left[\frac{2}{3} +R_j^2 -\frac{2\Delta^2}{3} +\frac{ \delta^2}{6} \right]
 \frac{K_j}{P_j^2}\right. \nonumber \\
&\left. \ \ \ \ \ \ \ \ \ \ \ +\frac{L_i}{4-R^2_{H_i}} \right\}
\\
A \left(\tpi\right):& \lambda=\pm 1 \ \ \  -4v\betb\lf g g_Z
 \left[-R_j+\frac{\delta^2}{2}
\right] \frac{K_j}{P_j^2}          \\
A \left( \tpii \right):&\lambda=\pm 1 \ \ \
4 v\betb g g_Z\left[-R_j +\frac{\delta^2}{2}\right] \frac{K_j}{P_j^2}
\\
&\lambda=0 \ \ \  -\frac{8}{\sqrt{3}}v\betb g g_Z\frac{K_j}{R_ZP_j^2}
\left\{1+R_j-\Delta^2 + \frac {\delta^2}{2} ( R_j-1) \right\}
\eeq\een
In eqs.(\ref{a14}) we have defined:
\ben\label{a16}\beq
\delta^2=&\frac{1}{2}(R_Z^2-R_P^2)\\
L_1=&\frac{1}{2} \sin (\alpha-\beta) T_{100},\ \ \ \
L_2= \frac{1}{2} \cos (\alpha-\beta) T_{200}\\
K_j=&O_{0j}^{''L}T_{P0j}
\eeq\een

\subsubsection* {5) $\chi \chi \rightarrow  W^-(\lambda)H^+ $ }
In this case one has contributions from the exchange of the two charginos,
which
are again labelled by the suffix $j$, as well as from the exchange of all three
neutral Higgs bosons. Notice also that there are equal contributions from
$W^+H^-$ and $W^-H^+$ production, leading to an overall factor of 2 in the
final
cross section in this case. The contributing partial wave amplitudes are:
\ben\label{a16a}\beq
A \left( \iso \right) :&\lambda=0\nonumber\\
&4\sqrt{2}\betb g^2\frac{\jp}{R_WP_j}
+ \sqrt{2}v^2\frac{ \betb g^2 (\jp)}{R_W P_j}
\left\{ \frac{5}{2} -\frac{2}{P_j}+\frac{4 \betb^2}{3P_j^2}
-\frac{2}{3P_j}\left[2-\Delta^2\right]
\right\}
\nonumber\\
&-4 \sqrt{2}\betb g^2(\jm) \frac{R_j^+}{R_WP_j}
-\sqrt{2}\betb v^2 g^2(\jm)\frac{R_j^+}{R_W P_j}
\left\{\frac{4 \betb^2}{3P_j^2}-\frac{2}{P_j}+\frac{1}{2}\right\}
\nonumber\\
&-4\sqrt{2} \betb g^2 T_{P00} \frac{1}{R_W}\frac{1}{4-R_P^2}
\left[1+\frac{3v^2}{8}\right]
\\
A \left(\tpo\right) :&\lambda=0\nonumber\\
&4\sqrt{6}\frac{v\betb}{R_W}g^2\left\{
R_j^+\left[1+\frac{\delta^2}{2}\right]
\frac{\km}{3P_j^2} - \left[\frac{2}{3}+R_j^2-\frac{2}{3}\Delta^2
+ \frac {\delta^2}{6} \right] \frac{\kp}{P_j^2}\right\}
\nonumber\\
&-\frac{4\sqrt{6}}{R_W}v\betb \frac{g^2L_i}{4-R_{H_i}^2}
\\
A \left(\tpi \right):&\lambda=\pm 1\nonumber\\
&-4v\left[{R_j^+}^2-\frac{\delta^4}{4}\right]g^2\frac{\jp}{P_j^2}
-4v g^2(\jm) \frac{R_j^+}{P_j}\nonumber\\
&+\frac{4v\betb \lambda}{P_j^2}
g^2\left\{ R_j^+(\km)-\frac{\delta^2}{2}(\kp)
\right\}\\
&\lambda=0 \ \ \ \ \ -\frac{4vg^2}{R_WP_j} \left[ (\jp)+(\jm) R_j\right]
\left\{1+\frac{\delta^2}{2} \right\}
\\
A \left(\tpii\right) : & \lambda=\pm 1\nonumber\\  &
-4v\lambda g^2\frac{\jp}{P_j^2} \overline{\beta}_f^2
+\frac{4v\betb }{P_j^2}g^2\left\{ (\kp)\frac{\delta^2}{2} -R_j^+(\km)\right\}
\\
&\lambda=0\nonumber\\
&\frac{8v \betb g^2}{\sqrt{3} R_W P_j^2} \left\{\left(-1+\Delta^2
+ \frac {\delta^2} {2} \right)(\kp)
-R_j^+ \left( 1 + \frac {\delta^2}{2} \right)(\km)\right\}
\eeq\een
In eqs(\ref{a16a}) we have introduced:
\ben\label{a17}\beq
\delta^2=&\frac{1}{2}(R_W^2-R_{H^+}^2)\\
F_j^{'}=& -\frac{1}{4} (\om)(\qp)\\
F_j^{''}=& -\frac{1}{4} (\op)(\qm)\\
K_j^{'}=&-\frac{1}{4}(\om)(\qm)\\
K_j^{''}=&-\frac{1}{4} (\op)(\qp)\\
\eeq\een

\subsubsection*{ 6) $\chi\chi \rightarrow  H_aH_b$ or $ PP$}
Both the $H_aH_b$ ($a,b = 1,2$) and the $PP$ final state can be produced by the
exchange of one of the four neutralinos (labelled by $j$) in the $t-$ or
$u-$channel, as well as by scalar Higgs exchange in the $s-$channel. The
contributing amplitudes for the $H_aH_b$ final state can be written as:
\ben\label{a19}\beq
A \left( \tpo \right):&
\sqrt{6} vg^2 \left\{ g_{abi}T_{i00} \frac{R_Z}{4-R_{H_i}^2+ i G_{H_i}}
-2 T_{a0j}T_{b0j} \frac{1+R_j}{P_j}
+\frac{4}{3}T_{a0j}T_{b0j} \frac{\betb^2}{P_j^2}\right\}
\\
A \left(\tpii \right):&
-\frac{8}{\sqrt{3}} v \betb^2 g^2T_{a0j} T_{b0j}\frac{1}{P_j^2}
\eeq\een
The corresponding $PP$ amplitudes can simply be obtained by replacing $a, b$
by $PP$ and  $R_j$ by $-R_j$. The rescaled width $G_{H_i} = \Gamma_{H_i}
m_{H_i} / m^2_{\chi}$ as before, and the trilinear Higgs couplings are
\cite{38}:
\ben\label{a19i}
\beq
g_{111}&=-\frac{3}{2\cw} \cos\!2\alpha\cos(\alpha+\beta)
\\
g_{222}&= -\frac{3}{2\cw}\cos\!2\alpha\sin(\alpha+\beta)
\\
g_{112}&=g_{121}=\frac{1}{2\cw}\left[2\sin\!2\alpha\cos(\alpha+\beta)
+\sin(\alpha+\beta)\cos\!2\alpha\right]
\\
g_{122}&=g_{221}=-\frac{1}{2\cw}\left[2\sin\!2\alpha\sin(\alpha+\beta)
-\cos(\alpha+\beta)\cos\!2\alpha\right]
\\
g_{PP1}&=g_{1PP}=\frac{\cos(\alpha+\beta)}{2 \cw}\cos\!2\beta
\\
g_{PP2}&=g_{2PP}=-\frac{\sin(\alpha+\beta)}{2 \cw}\cos\!2\beta
\eeq\een
Recall that the suffix 1 (2) refers to the heavier (lighter) Higgs scalar.
\subsubsection*{7) $\chi\chi \rightarrow   H_a P $}
This process receives contributions from diagrams where one of the four
neutralinos (labelled by $j$) is exchanged in the $t-$ or $u-$channel, as well
as from $s-$channel exchange of $Z$ or $P$ bosons. The result is:
\ben\label{a20}\beq
A \left(\iso \right):&\nonumber\\&
2\sqrt{2}g^2g_{aPP}T_{P00}\frac{R_Z} {4-R_P^2} \left( 1 + \frac {v^2}{8}
\right) + \sqrt{2}g^2g_{aPZ} \frac{R_P^2-R_a^2}{R_Z^2} (1-\frac{v^2}{8})
\nonumber\\
&
-4\sqrt{2}g^2 T_{a0j}T_{P0j} \frac{R_j}{P_j}
\left\{1+v^2\left[\frac{1}{8}-\frac{1}{2P_j} +\frac{\bar{\beta}_f^2}
{3P_j^2}\right]
\right\}
\nonumber\\
&
-\sqrt{2}(R_P^2-R_a^2)g^2T_{a0j}T_{P0j}
\left\{1+v^2\left[-\frac{1}{8}-\frac{1}{2P_j}+\frac{\bar{\beta}_f^2}
{3P_j^2}\right]
\right\}
\\
A \left( \tpi \right) :& -4v\betb\frac{g^2g_{aPZ}}
{4-R_Z^2 + i G_Z} -4v\betb g^2\frac{T_{a0i}T_{P0i}} {P_j}
\eeq\een
Here, $G_Z = M_Z \Gamma_Z/m^2_{\chi}$ as usual, and we have introduced the
combinations of couplings
\ben\label{a20i}\beq
g_{1PZ}&=-\frac{\sin(\alpha-\beta)}{2 \cos^2\!\theta_W}O_{00}^{''L}
\\
g_{2PZ}&=-\frac{\cos(\alpha-\beta)}{2 \cos^2\!\theta_W}O_{00}^{''L}
\eeq\een
\subsubsection*{8) $\chi\chi \rightarrow         H^+ H^- $}
This final state gets contributions from exchange of the two chargino
states, labelled by $j$, as well as from the exhange of the $Z$ boson and
the two scalar Higgs bosons, labelled by $i$. The nonvanishing amplitudes
are:
\ben\label{a21}\beq
A \left( \tpo \right) : &\ \ -\sqrt{6}vg^2\left\{ (\qlpr)
\left[\frac{1}{P_j}-\frac{2\overline{\beta}_f^2}{3P_j^2}
\right] +2\qlr\frac{R_j^+}{P_j} \right\}
\nonumber\\ &
\ \ +\sqrt{6}v g^2 R_W \frac{g_{+-i}}{4-R_{H_i}^2}
\\
A \left(\tpi \right):
&\ \  2v \betb g^2 (\qlmr)\frac{1}{P_j} + 4v \betb \frac{\cos
 2\theta_W}{\cos^2\theta_W} g^2 \frac{O^{''L}_{00}}{4-R_Z^2}
\\
A \left( \tpii\right):
&\ \ \frac{4}{\sqrt{3}}v \overline{\beta}_f^2 g^2(\qlpr)\frac{1}{P_j^2}
\eeq\een
In eqs.(\ref{a21}) we have introduced the quantities
\ben\label{a22}\beq
g_{+-1}&=-T_{100}\left[ \cos(\beta-\alpha)-\frac{\cos(\alpha+\beta)}
{2 \cos^2 \! \theta_W} \cos\!2\beta\right]\\
g_{+-2}&=-T_{200}\left[ \sin(\beta-\alpha)+\frac{\sin(\alpha+\beta)}
{2 \cos^2 \! \theta_W} \cos\!2\beta\right]
\eeq\een

\subsubsection*{ 9) $\chi\chi \rightarrow f_a(h)\bar{f}_a(\bar h) $}
Here we use the suffices $a,b$ to label the up and down components of weak
isodoublets, with $T_{3,a} = \pm 1/2$. This allows us to use the same notation
for quarks and leptons. Of course, in the final cross section a color factor of
3 has to be included in case of quarks.

This final state can be produced by the exchange of the two sfermion
eigenstates
in the $t-$ and $u-$channel, as well as by $s-$channel exchange of the $Z$
boson or of one of the three neutral Higgs bosons. Since the final state
particles can be massless, we have to include finite widths for all $s-$channel
propagators. We remind the reader that we do not expand $s-$channel propagators
with respect to $v$, while we do expand $t-$ and $u-$channel propagators. The
contributing partial wave amplitudes can be written as:
\ben\label{a23}\beq
A \left( \iso \right) :& \lambda_f=0
\nonumber\\
&\sqrt{2}\mbrh (\xpw)\left\{1+v^2
\left[-\frac{1}{2P_1}+\frac{\overline{\beta}_f^2}{3P_1^2} \right]\right\}
\frac{R_f}{P_1}
\nonumber\\
&+2\sqrt{2} \mbrh \xw
\frac{1}{P_1}\left\{1+v^2\left[\frac{1}{4}-\frac{1}{2P_1}
-\frac{\bar{\beta}_f^2}{6P_1}
+\frac{\overline{\beta}_f^2}{3P_1^2}\right]\right\}
\nonumber\\
&+( \xlrz,\wlry,\hlrl )
\nonumber\\
&+\mbrh\frac{2\sqrt{2}g^2}{\cos^2\theta_W}
O_{00}^{''L}T_{3a}\frac{R_f}{R_Z^2}
\nonumber\\
&+4\sqrt{2}\mbrh g h_{Pa}T_{P00}\frac{1}{4-R_P^2 + i G_P} \left( 1 + \frac
{v^2}{4} \right)
\\
A \left( \tpo \right) :&\lambda_f=0 \nonumber\\
&-\sqrt{6}v\betb (\xw)\left\{ \frac{1}{P_1}-\frac{2}{3P_1^2}\right\}
+\sqrt{6}v\betb (\xpw)\frac{R_f}{P_1^2}
\nonumber\\
&+( \xlrz,\wlry,\hlrl )
\nonumber\\
&-2\sqrt{6} v \betb g \frac{ h_{ia}T_{i00}}{4-R_{H_i}^2 + i G_{H_i} }
\\
A \left(\tpi \right) :&\lambda_f=0 \nonumber\\
&\frac{vR_f}{P_1}(\xmw)
+( \xlrz,\wlry,\hlrl )
\nonumber\\
&-2g^2 \frac{O_{00}^{''L}}{\cos^2\!\theta_W}
\left\{T_{3,a}-2e_{f_a}\sin^2 \! \theta_W\right\}\frac{R_f}{4-R_Z^2+iG_Z}
\\
&\lambda_f=\pm 1 \nonumber\\
&\sqrt{2} v \left\{\lambda_f \betb (\xpw)
\left[-\frac{1}{P_1}+\frac{1}{P_1^2}
   \right]\right.
\nonumber\\& \left.
+ (\xmw)\left[\frac{1}{P_1}-\frac{\bar{\beta}_f^2}{P_1^2}\right]
+2\betb \lambda_f ( \xw)\frac{R_f}{P_1^2}
\right\}
\nonumber\\
&+( \xlrz,\wlry,\hlrl )
\nonumber\\
&+2\sqrt{2}v g_Z^2 O_{00}^{''L}
\left\{ +\lambda_f T_{3,a}\betb - \left[T_{3,a}-2e_{f_a}
\sin^2\theta_W \right] \right\}\frac{1}{4-R_Z^2 + i G_Z}
\\
A \left( \tpii \right) :& \lambda_f=0\nonumber\\
&-\frac{2}{\sqrt{3}} v \betb \frac{1}{P_1^2}
\left\{R_f(\xpw)+2\xw\right\} \nonumber \\
&+( \xlrz,\wlry,\hlrl ) \\
&\lambda_f=\pm 1\nonumber\\
&\sqrt{2} v\betb \frac{1}{P_1^2}
\left\{- (\xpw) +\betb \lf(\xmw) -2  R_f \xw \right\}
 \nonumber \\ &+( \xlrz,\wlry,\hlrl )
\eeq\een
In eqs.(\ref{a23}), we have defined $R_f = m_{f_a}/\mchi$ as in
eq.(\ref{a5b3}).
These expressions fully include mixing between $SU(2)$ doublet and singlet
sfermions. The sfermion mass eigenstates, labelled by 1 and 2 in
eqs.(\ref{a23}), are defined by
\ben\label{a26b}
\beq
&\tilde{f}_1=\tilde{f}_L \cos\!\theta_f +\tilde{f}_R \sin\!\theta_f,\\
&\tilde{f}_2=-\tilde{f}_L \sin\!\theta_f +\tilde{f}_R \cos\!\theta_f;
 \eeq\een
the corresponding rescaled masses determine the inverse propagators $P_1$
and $P_2$ as shown in eq.(\ref{a5b1}). Sfermion mixing also affects their
couplings to neutralinos: \beq \label{anew}
&X_{a0}^{'}=X_{a0}\cos\!\theta_f+Z_{a0}\sin\!\theta_f,\ \ \
W_{a0}^{'}=Z_{a0}\cos\!\theta_f+Y_{a0}\sin\!\theta_f, \nonumber\\
&Z_{a0}^{'}=-X_{a0}\sin\!\theta_f+Z_{a0}\cos\!\theta_f,\ \ \
Y_{a0}^{'}=-Z_{a0}\sin\!\theta_f+Y_{a0}\cos\!\theta_f. \eeq
The couplings of the unmixed sfermions in eq.(\ref{anew}) can be found
in ref.\cite{38}\footnote{Notice that the last term in the relevant eq.(5.5)
of that reference has a wrong sign; H. Haber, private communication.}:

\ben\label{a25}\beq
&X_{a0}=-\sqrt{2} g \left[T_{3a} N_{02}
-\tan\theta_W(T_{3a}-e_{f_a})N_{01}
\right]
\\
&Y_{a0}=\sqrt{2} g \tan\theta_W e_{f_a} N_{01}
\\
&Z_{u0}=-\frac{gm_u}{\sqrt{2} \sin\!\beta M_W}N_{03}
\nonumber\\
&Z_{d0}=-\frac{gm_d}{\sqrt{2} \cos\!\beta M_W}N_{04}
\eeq\een

Finally, the couplings between Higgs bosons and SM fermions are \cite{38}:

\ben\label{a26}\beq
&h_{Pu}=-\frac{gm_u\cot\!\beta}{2M_W}
,\ \ \ h_{Pd}=-\frac{gm_d\tan\!\beta}{2M_W}
\\
&h_{1u}=-\frac{gm_u\sin\!\alpha}{2M_W\sin\!\beta}
,\ \ \ h_{1d}=-\frac{gm_d\cos\!\alpha}{2M_W\cos\!\beta}
\\
&h_{2u}=-\frac{gm_u\cos\!\alpha}{2M_W\sin\!\beta}
,\ \ \ h_{2d}=\frac{gm_d\sin\!\alpha}{2M_W\cos\!\beta}
\eeq\een

\setcounter{footnote}{0}
\section*{Appendix B: Application of the Equivalence Theorem}
As already discussed in sec. 2b, the high--energy limit of the amplitudes for
the production of longitudinal gauge bosons can be understood from the
equivalence theorem \cite{46}. This also provides a useful check of these
amplitudes. As an example, we discuss in this Appendix the production of two
longitudinal $Z$ bosons. We see from eqs.(\ref{a9}) that only two amplitudes
contribute to the $\lambda = \bar{\lambda} = 0$ final state; in the limit
$|\mchi| \gg M_Z$ they become
\ben \label{b1} \beq
A \left( \tpo \right) &= - \frac { 4 \sqrt{6} g_Z^2 v} {R_Z^2 P_j}
{O_{0j}^{''L}}^2 \left[ \frac{2}{3} + R_j \left( 1 - \frac {2} {3 P_j}
\right) \right] + \frac {2 \sqrt{6} v g_2 F_i} {\left( 4 - R^2_{H_i}
\right) R_W }; \label{b1a} \\
A \left( \tpii \right) &= - \frac {8 v g_Z^2} {\sqrt{3} R_Z^2 P_j}
{O_{0j}^{''L}}^2 \left( 1 + \frac {2 R_j} {P_j} \right) , \label{b1b} \eeq \een
where we have already made use of the fact that ${O_{0j}^{''L}}^2 \propto
R_Z^2$, so that only contributions with an explicit factor $1/R_Z^2$ survive
in the high--energy limit. More specifically, we see from eqs.(\ref{e6}) that
only the contributions from the higgsino--like neutralinos ($j=3,4$) will
survive in this limit:
\ben \label{b2} \beq
R_3 &= - R_4 = \frac {\mu} {M_1}; \label{b2a} \\
P_3 &= P_4 = 1 + \frac {\mu^2} {M_1^2}; \label{b2b} \\
{O_{03}^{''L}}^2 + {O_{04}^{''L}}^2 &=  \frac{1}{4}
\left( \frac {M_Z \sin \! \theta_W} {M_1^2 - \mu^2} \right)^2 \left( M_1^2
+ \mu^2 + 2 M_1 \mu \sin \! 2 \beta \right); \label{b2c} \\
{O_{03}^{''L}}^2 - {O_{04}^{''L}}^2 &= - \frac{1}{4}
\left( \frac {M_Z \sin \! \theta_W} {M_1^2 - \mu^2} \right)^2 \left[
\left( M_1^2 + \mu^2 \right) \sin \! 2 \beta + 2 M_1 \mu \right]. \label{b2d}
\eeq \een
In SUGRA, $m_{\chi}^2 \gg M^2_Z$ almost always implies $m_P^2 \gg M_Z^2$.
In that case $F_1$ becomes very small, so that the contribution from the heavy
Higgs scalar can be neglected; moreover, one has $R^2_{H_2} \ll 1$ in this
limit, and $\alpha = \beta + \pi/2$. This gives:
\be \label{b3}
F_2 = - T_{200} = \frac {M_Z \sin^2 \! \theta_W} {\cos \! \theta_W \left(
\mu^2 - M_1^2 \right) } \left( M_1 + \mu \sin \! 2 \beta \right), \ee
where we have used eq.(\ref{e6a}). Inserting eqs.(\ref{b2}) and (\ref{b3})
into (\ref{b1}) finally yields:
\ben \label{b4} \beq
A \left( \tpo \right) &= \sqrt{6} v g'^2 \frac {M_1^2} {\mu^2 + M_1^2}
\frac {1} { \left( \mu^2 - M_1^2 \right)^2} \left\{ \frac{2}{3} M_1^2 \left(
M_1^2 + \mu^2 + 2 M_1 \mu \sin \! 2 \beta \right) \right.
\nonumber \\
&\left. \hspace*{3cm}
- M_1 \mu \left( 1 - \frac{2}{3} \frac {M_1^2} {M_1^2 + \mu^2}
\right) \left[ \left( M_1^2 + \mu^2 \right) \sin \! 2 \beta + 2 M_1 \mu \right]
\right\}  \nonumber \\
&+ \frac {\sqrt{6}}{2} g'^2 v \frac {M_1} {\mu^2 - M_1^2} \left( M_1 + \mu
\sin \! 2 \beta \right); \label{b4a} \\
A \left( \tpii \right) &= - \frac {2} {\sqrt{3}} v g'^2 \frac {1}
{M_1^2 + \mu^2} \left( \frac {M_1^2} {M_1^2-\mu^2} \right)^2 \left\{
M_1^2 + \mu^2 + 2 M_1 \mu \sin \! 2 \beta \right. \nonumber \\ & \left.
\hspace*{3cm}
 - \frac {2 \mu M_1} {M_1^2 + \mu^2} \left[ \left( M_1^2 + \mu^2 \right)
\sin \! 2 \beta + 2 M_1 \mu \right] \right\}. \eeq \een
The last term in eq.(\ref{b4a}) comes from Higgs exchange. One recognizes the
form of our ``symbolic'' expression (\ref{e12}) of sec. 2b.

On the other hand, since the neutral Goldstone boson $G$ is a pseudoscalar,
the amplitudes for the production of a $GG$ pair
can be directly read off from our results for $PP$ production,
eqs.(\ref{a19}).
Retaining only those terms that remain finite as $m^2_{\chi} / M^2_Z \to
\infty$, one has \ben \label{b5} \beq
A \left( \tpo \right) &= - \frac {2 \sqrt{6} v g^2} {P_j} {T_{G0j}}^2
\left( 1 - R_j - \frac {2} {3 P_j} \right); \label{b5a} \\
A \left( \tpii \right) &= - \frac {8} {\sqrt{3}} v g^2 {T_{G0j}}^2 \frac {1}
{P^2_j} . \label{b5b} \eeq \een
The couplings $T_{G0j}$ can be obtained from eq.(\ref{a5a1}) by the
substitution
$\sin \! \beta \to - \cos \! \beta, \ \cos \! \beta \to \sin \! \beta$.
We see that again only the contributions
from the exchange of the higgsino--like neutralinos ($j=3,4$) survive:
\ben \label{b6} \beq
{T_{G03}}^2 + {T_{G04}}^2 &= \frac {1} {4} \tan^2 \! \theta_W; \label{b6a} \\
{T_{G03}}^2 - {T_{G04}}^2 &= - \frac {1} {4} \tan^2 \! \theta_W
\sin \! 2 \beta. \label{b6b} \eeq \een
Inserting this into eqs.(\ref{b5}) gives: \ben \label{b7} \beq
A \left( \tpo \right) &= - \frac {\sqrt{6}}{2} v g'^2
\frac {M_1^2} {M_1^2+\mu^2} \left( 1 - \frac{2}{3} \frac {M_1^2} {M_1^2 +
\mu^2}
+ \frac {\mu}{M_1} \sin \! 2 \beta \right) ; \label{b7a} \\
A \left( \tpii \right) &= - \frac {2}{\sqrt{3}} v g'^2 \left( \frac {M_1^2}
{M_1^2 + \mu^2} \right)^2. \label{b7b} \eeq \een
While perhaps not immediately obvious, it can be shown that eqs.(\ref{b7})
are indeed identical to eqs.(\ref{b4})\footnote{Up to an overall sign between
the $\tpo$ amplitudes; however, as shown in the Appendix of ref.\cite{6}, the
signs of noninterfering amplitudes are ambiguous when Majorana spinors are
involved.}.

The fact that the Goldstone amplitudes (\ref{b7}) automatically yield more
compact expressions shows that they can not only be used to check some
amplitudes of Appendix A, but they also allow to derive the high--energy
limit of the amplitudes for the production of longitudinal gauge bosons more
quickly. We saw already that the amplitudes for the production of
transverse gauge bosons vanish with some power of $M_Z/\mchi$ in the limit of
large $|\mchi|$ if $X$ is bino--like; the longitudinal helicity
states therefore dominate the total production of gauge bosons in this
important special case. (This is not true for higgsino--like or mixed LSP,
however.) Most amplitudes for the production of Goldstone bosons can be
read off directly from our results for the production of pseudoscalar and
charged Higgs bosons, replacing
$\sin \! \beta \to - \cos \! \beta, \ \cos \! \beta \to \sin \! \beta$
in the
corresponding couplings of Higgs bosons to neutralinos and charginos.
The only exception is $H^+ G^-$ production, which gives the high--energy limit
of $H^+ W^-_L$ production. We list the nonvanishing partial wave amplitudes
for completeness:
\ben\label{a18}\beq
\iso: &\ \  2\sqrt{2}g^2\qlr\frac{R_j^+}{\pjp}\left\{
1+v^2\left[\frac{1}{8} -\frac{1}{2(\pjp)} +\frac{1}{3(\pjp)^2}
\right]\right\}
\\
\tpo:&\nonumber\\
&\ \ -\frac{\sqrt{6}}{2} v \sin\!2\beta g^2(\qlmr)\left\{\frac{1}{\pjp}
-\frac{2}{3(\pjp)^2} \right\}
\nonumber\\
&\ \ \ \ \ \ \ \ \ \ \ +\sqrt{6}v g^2\qlr \cos\!2\beta\frac{R_j^+}
{\pjp}
\\
\tpi:&\ \ v\sin\!{2\beta} g^2({Q^{'L}_{ij}}^2+{Q^{'R}_{ij}}^2)\frac{1}{\pjp}
\\
\tpii:&\ \  -\frac{2}{\sqrt{3}}v\sin\!{2\beta}g^2\frac{\qlmr}{(\pjp)^2}
\eeq\een
In eqs.(\ref{a18}), we have again only kept terms that remain finite at
high energies.

\newpage
\section*{Figure Captions}

\renewcommand{\labelenumi}{Fig. \arabic{enumi}}
\begin{enumerate}
\item  
The rescaled LSP relic density \oh\ as a function of the LSP mass \mchi\ for
scalar SUSY breaking mass parameter $m$ = 300 GeV and four combinations of the
ratio \tanb\ of Higgs vevs and the mass $m_t$ of the top quark. The solid
(dashed) curves are for positive (negative) gaugino mass parameter $M$. Notice
that $M$, the trilinear soft breaking parameter $A$ and the supersymmetric
Higgs(ino) mass parameter $\mu$ all vary along the curves, due to the relations
between model parameters implied by radiative gauge symmetry breaking; in
particular, the locations of the pole and thresholds caused by the light scalar
Higgs boson are different in each case.

\vspace*{5mm}

\item    
An enlargement of the threshold region of the solid curves in figs. 1a,b. For
$m_t$ = 130 GeV (solid curve) the $hh$ and $Zh$ final states are accessible
for $\mchi >$ 75 and 84 GeV, respectively, while for $m_t$ = 160 GeV (dashed),
they are only accessible for $\mchi >$ 95 and 93 GeV, respectively.

\vspace*{5mm}

\item    
The rescaled LSP density as a function of the LSP mass for two examples of
higgsino--like LSPs. $M, \ m$ and $m_t$ are fixed for each curve, while
$\mu, \ A$ and \tanb\ vary. For $M=-1$ TeV (dashed curve) the LSP becomes
bino--like at $\mchi \simeq 430$ GeV, which explains the rapid rise of
this curve at large \mchi. Only $\chi \chi$ annihilation has been included
in the calculation; this
substantially overestimates \oh\ for $\mchi \leq M_W$, as argued in the
text.

\vspace*{5mm}

\item    
Demonstration of the importance of the SUGRA imposed relations between particle
masses. The solid curve shows the full SUGRA prediction for \oh\ as a function
of \tanb; eq.(\ref{e23}) implies that $\mchi = m_P/2$ at \tanb=35.3, where $P$
denotes the pseudoscalar Higgs boson. The long dashed curve has been obtained
by keeping $m_P$ constant at the value SUGRA predicts for \tanb=2, i.e.
$m_P=783$ GeV, but the $\tilde{\tau}$ and $\tilde{b}$ masses are still allowed
to vary with \tanb. For the short dashed curve all effects of the $b$ and
$\tau$ Yukawa couplings have been switched off.

\vspace*{5mm}

\item     
Contours of constant \oh\ in the plane spanned by $M$ and \tanb, for fixed
$m=250$ GeV and $m_t$ = 140 GeV; results for all 4 combinations of signs of $M$
and \tanb\ are shown. The solid and long dashed lines are contours where \oh=1
and 0.25, respectively. The short dashed lines in (a) are contours where
\oh=0.025; since these contours are very close to the $Z, \ h$ or $P$ poles,
in (b) - (d) we have merely indicated the location of these poles by the
short dashed lines. The region outside of the dotted curves is exlcuded by
various experimental and theoretical constraints, as described in the text.
A flat universe requires $0.25 \leq \oh \leq 1$, while LSPs can build up the
dark matter halo of galaxies if $\oh \geq 0.025$.

\vspace*{5mm}

\item   
Contours of constant \oh\ in the plane spanned by $M$ and \tanb, for the
quadrant $M>0, \ \tanb>0$, and four different combinations of $m$ and $m_t$
as indicated. The notation is as in figs. 5 (b)--(d).

\vspace*{5mm}

\item  
The upper bound on $|M|$ which follows from the
requirement that LSPs do not overclose the universe, $\oh \leq 1$. The
long dashed, solid, and short dashed curves are valid for $m_t$ = 110, 140
and 170 GeV, respectively, and show the absolute upper bound over the entire
experimentally and theoretically allowed parameter space. The dotted curves
emerge when one restricts the parameter space to $|\tanb| \leq 15$; the upper
(lower) dotted curve is for $m_t$ = 140 (170) GeV. In all cases the maximal
$|M|$ is found in the quadrant $M<0, \ \tanb<0$.

\vspace*{5mm}

\item    
Upper bounds on \oh\ as a function of the mass \mlr\ of the $SU(2)$ singlet
sleptons. The long dashed, solid, and short dashed curves are for $m_t$ = 110,
140 and 170 GeV, respectively, and present
experimental bounds on sparticle masses; the long-short dashed curve is
valid if the lower bounds on the masses of charged sparticles and gluinos are
raised to 80 and 200 GeV, respectively, for $m_t$ = 140 GeV. The dotted curve
is based on a simple analytical approximation that assumes that the LSP is a
pure bino, as described in the text.

\end{enumerate}

\end{document}